\documentclass[graybox]{svmult}

\usepackage{mathptmx}       
\usepackage{helvet}         
\usepackage{courier}        
\usepackage{type1cm}        
\usepackage{makeidx}         
\usepackage{graphicx}        
\usepackage{multicol}        
\usepackage[bottom]{footmisc}

\usepackage{amssymb, amsmath}
\usepackage{subfigure}
\usepackage{url}

\usepackage{booktabs}

\makeindex             

\begin{document}

\title*{Computers from plants we never made. Speculations.}
\author{Andrew Adamatzky \and Simon Harding \and Victor Erokhin \and Richard Mayne
\and Nina Gizzie \and Frantisek Baluska \and Stefano Mancuso \and Georgios Sirakoulis}

\authorrunning{Adamatzky, Harding, Erokhin, Mayne, Gizzie, Baluska, Mancuso, Sirakoulis} 

\institute{
Andrew Adamatzky \and Simon Harding \and Richard Mayne \and Nina Gizzie
\at 
Unconventional Computing Centre, UWE, Bristol, United Kingdom,
\email{andrew.adamatzky@uwe.ac.uk}, \email{slh@evolutioninmaterio.com},
\email{richard.mayne@uwe.ac.uk}, \email{nina.gizzie@gmail.com}
\and 
Victor Erokhin 
\at
CNR-IMEM, Parma, Italy, \email{victor.erokhin@fis.unipr.it}
\and 
Frantisek Baluska
\at 
Institute of Cellular and Molecular Botany, University of Bonn, Germany,
\email{unb15e@uni-bonn.de}
\and 
Stefano Mancuso 
\at 
International Laboratory of Plant Neurobiology, University of Florence, Italy,
\email{stefano.mancuso@unifi.it}
\and 
Georgios Ch. Sirakoulis
\at 
Department of Electrical \& Computer Engineering,  
Democritus University of Thrace, Xanthi, Greece,
\email{gsirak@ee.duth.gr}
}


\maketitle

\abstract*{Plants are highly intelligent organisms. They continuously make distributed processing of sensory information, concurrent decision making and parallel actuation. The plants are efficient green computers per se. Outside in nature, the plants are programmed and hardwired to perform a narrow range of tasks aimed to maximize the plants' ecological distribution, survival and reproduction. To `persuade' plants to solve tasks outside their usual range of activities, we must either choose problem domains which homomorphic to the plants natural domains or modify biophysical properties of plants to make them organic electronic devices. We discuss possible designs and prototypes of computing systems that could be based on morphological development of roots, interaction of roots, and analog electrical computation with plants, and plant-derived electronic components. In morphological plant processors data are represented by initial configuration of roots and configurations of sources of attractants and repellents; results of computation are represented by topology of the roots' network. Computation is implemented by the roots following gradients of attractants and repellents, as well as interacting with each other. Problems solvable by plant roots, in principle, include shortest-path, minimum spanning tree, Voronoi diagram, $\alpha$-shapes, convex subdivision of concave polygons. Electrical properties of plants can be modified by loading the plants with functional nanoparticles or coating parts of plants of conductive polymers. Thus, we are in position to make living variable resistors, capacitors, operational amplifiers, multipliers, potentiometers and fixed-function generators. The electrically modified plants can implement summation, integration with respect to time, inversion, multiplication, exponentiation, logarithm, division. Mathematical and engineering problems to be solved can be represented in plant root networks of resistive or reaction elements.  Developments in plant-based computing architectures will trigger emergence of a unique community of biologists, electronic engineering and computer scientists working together to produce living electronic devices which future green computers will be made of.}

\abstract{Plants are highly intelligent organisms. They continuously make distributed processing of sensory information, concurrent decision making and parallel actuation. The plants are efficient green computers per se. Outside in nature, the plants are programmed and hardwired to perform a narrow range of tasks aimed to maximize the plants' ecological distribution, survival and reproduction. To `persuade' plants to solve tasks outside their usual range of activities, we must either choose problem domains which homomorphic to the plants natural domains or modify biophysical properties of plants to make them organic electronic devices. We discuss possible designs and prototypes of computing systems that could be based on morphological development of roots, interaction of roots, and analog electrical computation with plants, and plant-derived electronic components. In morphological plant processors data are represented by initial configuration of roots and configurations of sources of attractants and repellents; results of computation are represented by topology of the roots' network. Computation is implemented by the roots following gradients of attractants and repellents, as well as interacting with each other. Problems solvable by plant roots, in principle, include shortest-path, minimum spanning tree, Voronoi diagram, $\alpha$-shapes, convex subdivision of concave polygons. Electrical properties of plants can be modified by loading the plants with functional nanoparticles or coating parts of plants of conductive polymers. Thus, we are in position to make living variable resistors, capacitors, operational amplifiers, multipliers, potentiometers and fixed-function generators. The electrically modified plants can implement summation, integration with respect to time, inversion, multiplication, exponentiation, logarithm, division. Mathematical and engineering problems to be solved can be represented in plant root networks of resistive  or reaction elements.   Developments in plant-based computing architectures will trigger emergence of a unique community of biologists, electronic engineering and computer scientists working together to produce living electronic devices which future green computers will be made of.}

\section{Introduction}
\label{introduction}

Plants perform neuronal-like computation not just for rapid and effective adaptation to an ever-changing physical environment but also for the sharing of information with other plants of the same species and for communication with bacteria and fungi~\cite{baluvska2006neurobiological, baluvska2004root, baluvska2005plant, baluvska2009plant, baluvska2009deep, baluvska2010root, baluvska2006communication, baluvska2016vision, brenner2006plant}. In fact, plants emerge as social organisms. Roots are well known for their ability to avoid dangerous places by actively growing away from hostile soil patches. By using a vast diversity of volatiles, plants are able to attract or repel diverse insects and animals, as well as to shape their biotic niche. The number of volatile compounds released and received for plant communication is immense, requiring complex signal-release machinery, and 'neuronal-like' decoding apparatus for correct interpretation of received signals. The plant screens continuously and updates diverse information about its surroundings, combines this with the internal information about its internal state and makes adaptive decisions that reconcile its well-being with the environment~\cite{trewavas2005green, trewavas2009plant, trewavas2011ubiquity}. Until now, detection and contextual filtering of at least 20 biological, physical, chemical and electrical signals was documented. Plants discriminate lengths, directions and intensities of the signal. These signals induce a memory that can last, depending on the signal, for hours or days, or even up to years. Once learnt, these plant memories usually ensure a much quicker and more forceful response to subsequent signalling~\cite{plantsignallingbook, trewavas2009plant}. Via feedback cross-talks, memories become associative, inducing specific cellular changes. Thus plant behaviour is active, purpose-driven and intentional. 

In last decade we manufactured a range of high-impact experimental prototypes from unusual substrates. These include experimental implementations of logical gates, circuits and binary adders employing interaction of wave-fragments in light-sensitive Belousov-Zhabotinsky media~\cite{costello2005experimental, adamatzky2011computing, costello2011towards}, swarms of soldier crabs~\cite{gunji2011robust},
growing lamellypodia of slime mould {\emph Physarum polycephalum}~\cite{adamatzky2016advances},
crystallisation patterns in `hot ice'~\cite{adamatzky2009hot},
jet streams in fluidic devices~\cite{morgan2016simple}. After constructing over 40 computing, sensing and actuating devices with slime mould~\cite{adamatzky2016advances} we turned our prying eyes to plants.
We thought that plant roots could be ideal candidates to make unconventional computing because of the following features. Roots interact with each of their proximal neighbours and can be influenced by their neighbours to induce a tendency to align the directions of their growth~\cite{baluvska2010root, masi2009spatiotemporal, ciszak2012swarming}. Roots, as already mentioned, communicate their sensory information within and between adjacent roots, and other organisms like bacteria and fungi~\cite{bais2004plants, sugiyama2012root}. This signalling and communicative nature of plant roots is very important for our major task to generate root-based computing devices. Outside in Nature, almost all roots are interconnected into huge underground root-fungal information networks, resembling the Internet.
Roots can be grown in wet air chambers, as well as within nutrition solutes which allow accomplishment of relevant experiments. Roots can be isolated from seedlings using root vitro cultures which can be maintained for several years.
Each root apex acts both as a sensory organ and as a brain-like command centre to generate each unique plant/root-specific cognition and behaviour~\cite{baluvska2005plant, baluvska2004root, baluvska2009plant}. It is easy to manipulate root behaviour using diverse physical and chemical cues and signals. Roots efficiently outsource external computation by the environment via the sensing and suppression of nutrient gradients. The induction of pattern type of roots behaviour is determined by the environment, specifically nutrient quality and substrate hardness, dryness etc. Moreover, roots are sensitive to illumination and show negative phototropism behaviour when exposed to light~\cite{yokawa2011illumination, burbach2012photophobic}, electric fields, temperature gradients, diverse molecules including volatiles and, therefore, allow for parallel and non-destructive input of information.

In 2012 Adamatzky, Baluska, Mancuso and Erokhin submitted ERC grant proposal ``Rhizomes''. They proposed to experimentally realize computing schemes via interaction of plant roots and to produce electronic devices from living roots by coating and loading the roots with functional materials, especially conductive polymers. This particular grant proposal was not funded however some ERC evaluators `borrowed' methods and techniques from the proposal and implemented them in their own laboratories. Most notable example is plant electronics ideas: they were originally outlined in  ``Rhizomes'' proposal yet borrowed from the proposal by some Nordic fellows and published without crediting original authors.  Realising that most ideas will be `stolen' in a similar fashion sooner or later anyway, we decided to disclose them in full.

\section{Morphological computation}
\label{Morphologicalcomputation}

In morphological phyto-computers, data are represented by initial configuration of roots and configurations of sources of attractants and repellents. Results of computation are represented by the topology of the roots' network. Computation is implemented by the roots following gradients of attractants and repellents and interacting with each other.  Gradients of fields generated by configurations and concentrations of attractants are an important prerequisite for successful programming of plant-based morphological processors. Chemical control and programming of root behaviour can be implemented via application of attractants and repellents in the roots' growth substrate and/or in the surrounding air. In particular, chemo-attractants are carbohydrates, peptones, amino-acids phenylalanine, leucine, serine, asparagine, glycine, alanine, aspartate, glutamate, threonine, and fructose, while chemo-repellents are sucrose, tryptophan and salt. Plant roots perform complex computations by the following general mechanisms: root tropisms stimulated via attracting and repelling spatially extended stimuli; morphological adaptation of root system architecture to attracting and repelling spatially extended stimuli; wave-like propagation of information along the root bodies via plant-synapse networks; helical nutation of growing root apexes, patterns of which correlate with their environmental stimuli; competition and entrainment of oscillations in their bodies.

\subsection{Shortest path}
\label{shortestpath}

Shortest path \index{shortest path} is a typical problem that every rookie in computer science solves, typically by coding the treasured Knuth algorithm. 
In a less formal definition, given a space with obstacles, source and destination points, the problem is to calculate a collision-free path from the source to the destination which is the shortest possible relative to a predefined distance metric. A root is positioned at the source, the destination is labelled with chemo-attractants, obstacles are represented by source of chemo-repellents and/or localised super-threshold illumination. The root-based search could be implemented without chemicals but the source and destination loci are represented by poles of AC/DC voltage/current and the obstacles by domains of super-threshold illumination or sub-threshold humidity. The collision-free shortest path is represented by the physical body of the root grown from the source to the destination. Implementation of this task could require suppression of the roots branching. 
 
 \begin{figure}[!tbp]
\centering
\subfigure[]{\includegraphics[width=0.4\textwidth]{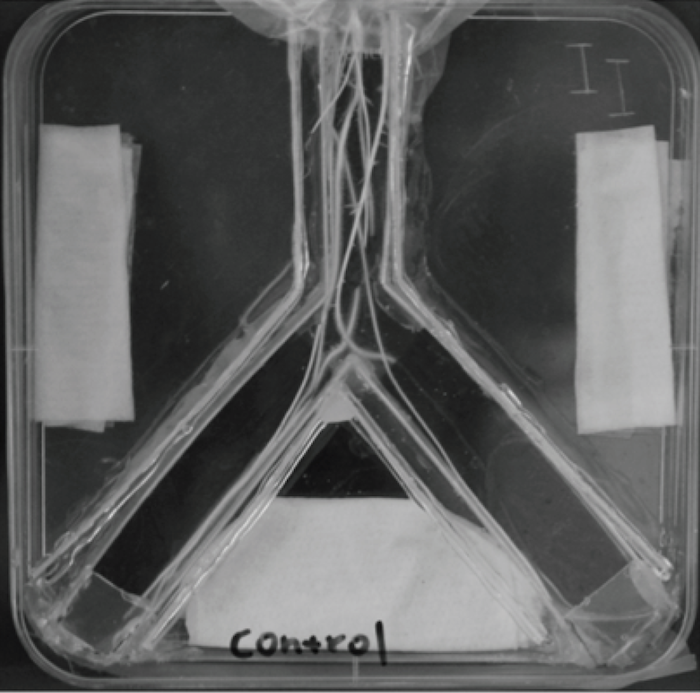}}
\subfigure[]{\includegraphics[width=0.4\textwidth]{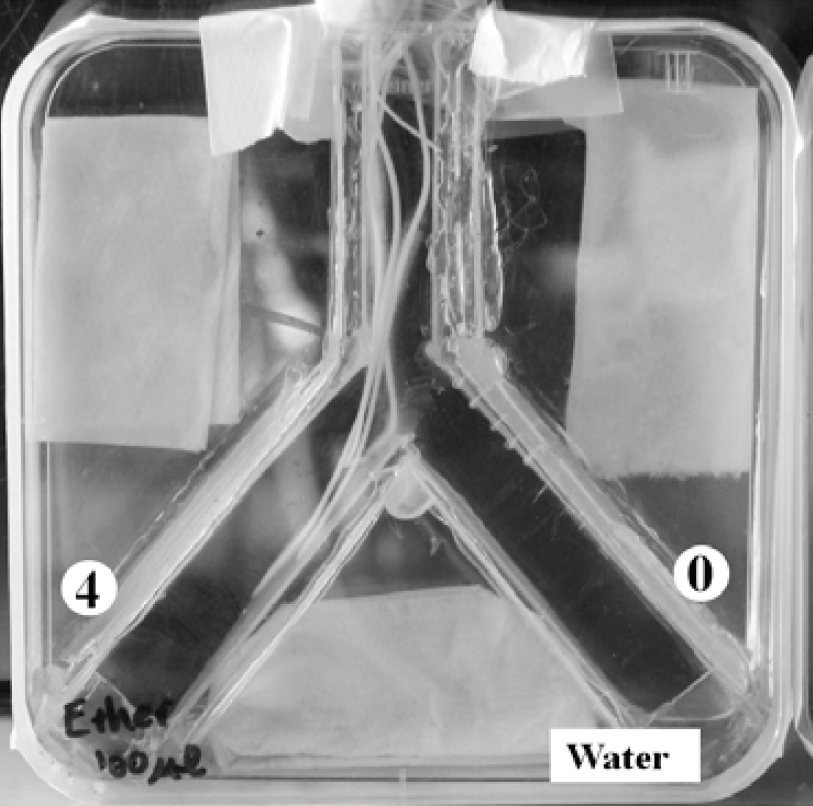}}
\caption{Plant roots select a path towards attractant. Reprinted from \cite{yokawa2014binary}.
(a)~Control experiment. The maize roots are growing Y-maze with 1~ml of distilled water in both slanted channels. 
(b)~Diethyl ether (100 $\mu$l ether and 900 $\mu$l distilled water is added to left slanted channel.}
\label{baluskamaze}
\end{figure}

\begin{figure}[!tbp]
\centering
\subfigure[]{\includegraphics[width=0.4\textwidth]{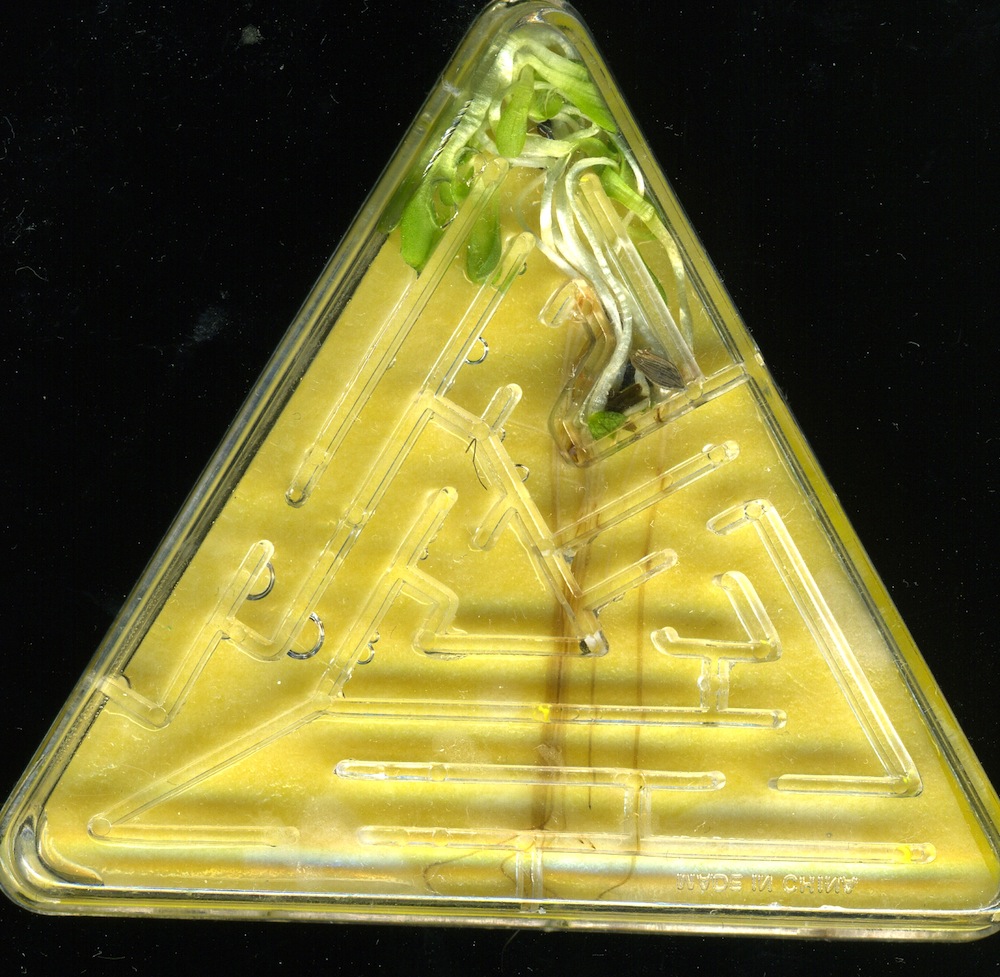}}
\subfigure[]{\includegraphics[width=0.4\textwidth]{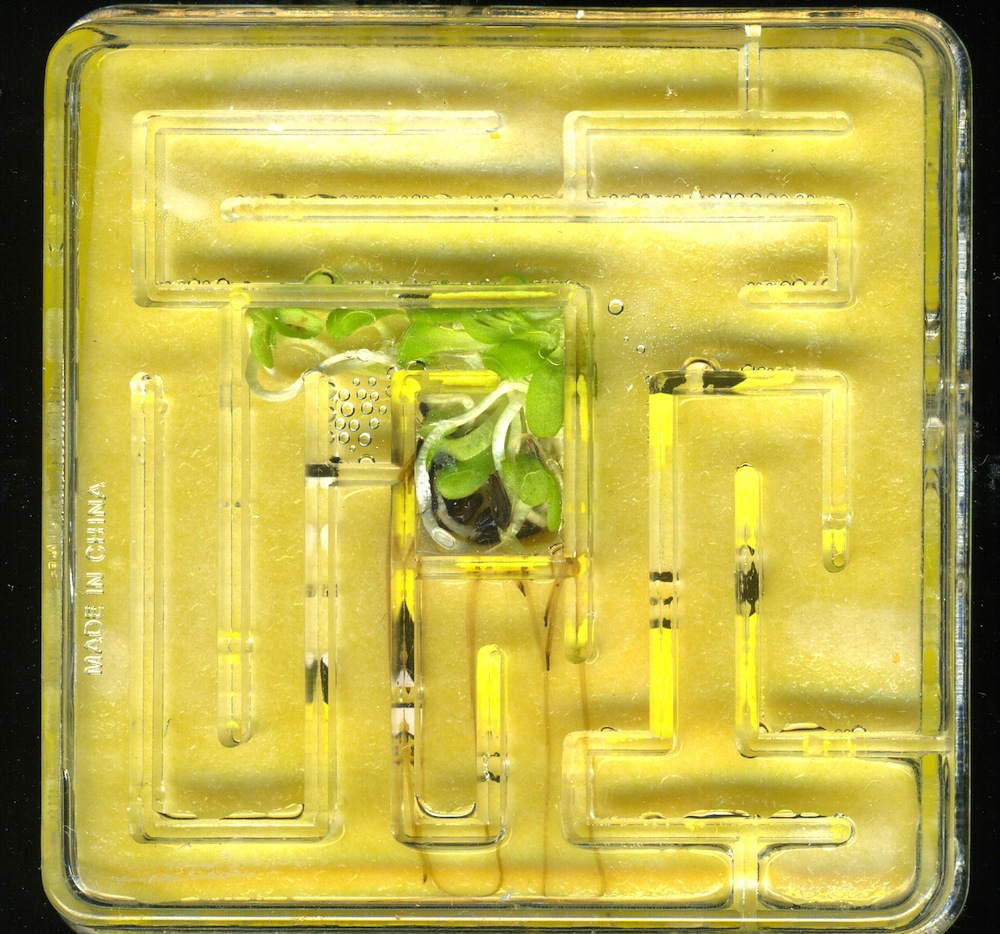}}
\caption{Scoping experiments on routing plant roots in mazes. From \cite{adamatzky2014towards}.
}
\label{adamatzkymazes}
\end{figure}

An experimental laboratory demonstration for showing feasibility of solving the shortest path with roots was demonstrated in Baluska lab and published in \cite{yokawa2014binary}. Their experiments have been done in Y-junction, turned upside down because roots are guided by gravity. Seeds were placed in the end of the vertical channel and attracting or neutral substances in the slanted channels. When just a distilled water is placed in both slanted channels, the roots from several seeds split arbitrarily between the channels (Fig.~\ref{baluskamaze}a). When a chemo-attractant, \index{chemo-attractant} e.g. diethyl ether, is placed in one of the slanted channels all roots move into the channel with the chemoattractant (Fig.~\ref{baluskamaze}b). 

Experiments on path finding by roots in mazes with more complex geometry than Y-junction are proved to be inconclusive so far. In \cite{adamatzky2014towards} we presented results of few scoping experiments, see two illustrations in Fig.~\ref{adamatzkymazes}, yet reached no conclusive results. When seeds are placed in or near a central chamber of a
labyrinth their routes somewhat grow towards exit of the labyrinth. However, they often become stuck midway and do not actually reach the exit. These findings have been confirmed also with roots of Arabidopsis (unpublished data). The inconclusive results are possibly due to the fact that we used gravity as the only guiding force to navigate the routes.  
\begin{figure}[!tbp]
\centering
\subfigure[]{\includegraphics[width=0.48\textwidth]{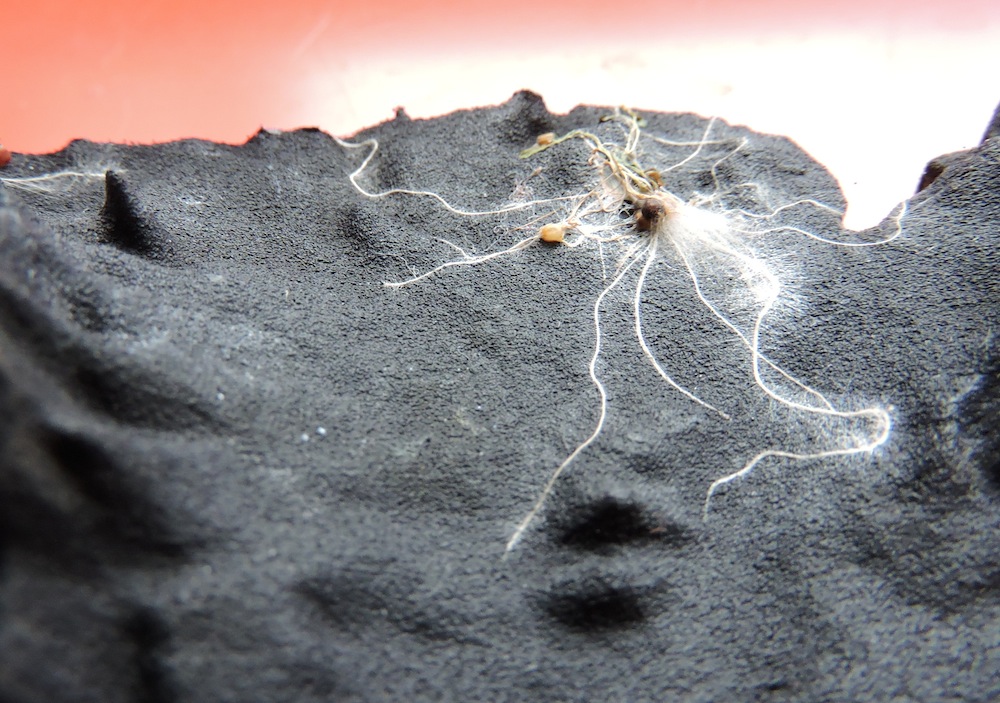}}
\subfigure[]{\includegraphics[width=0.48\textwidth]{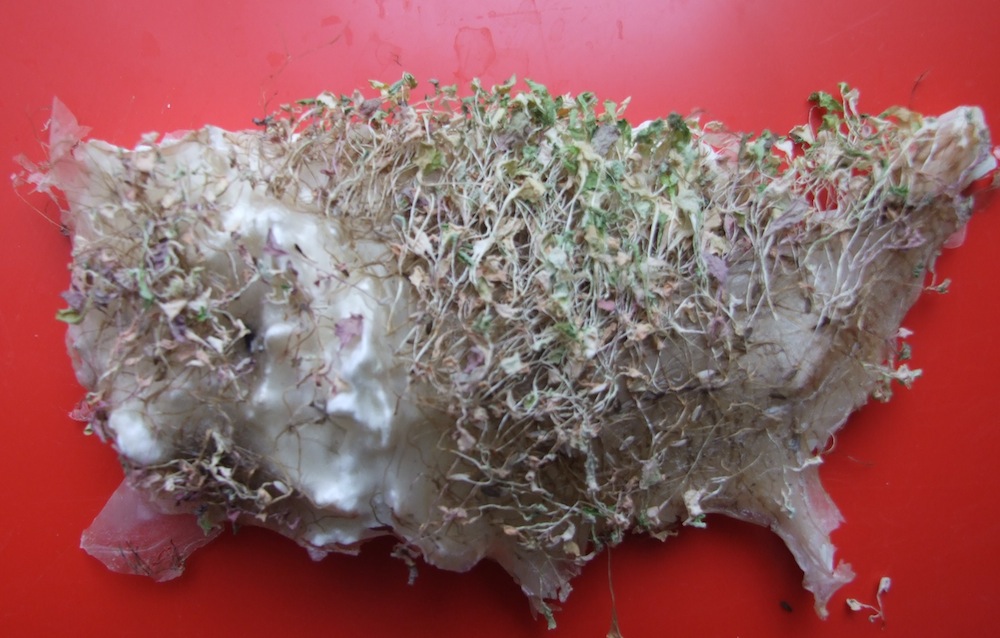}}
\subfigure[]{\includegraphics[width=0.48\textwidth]{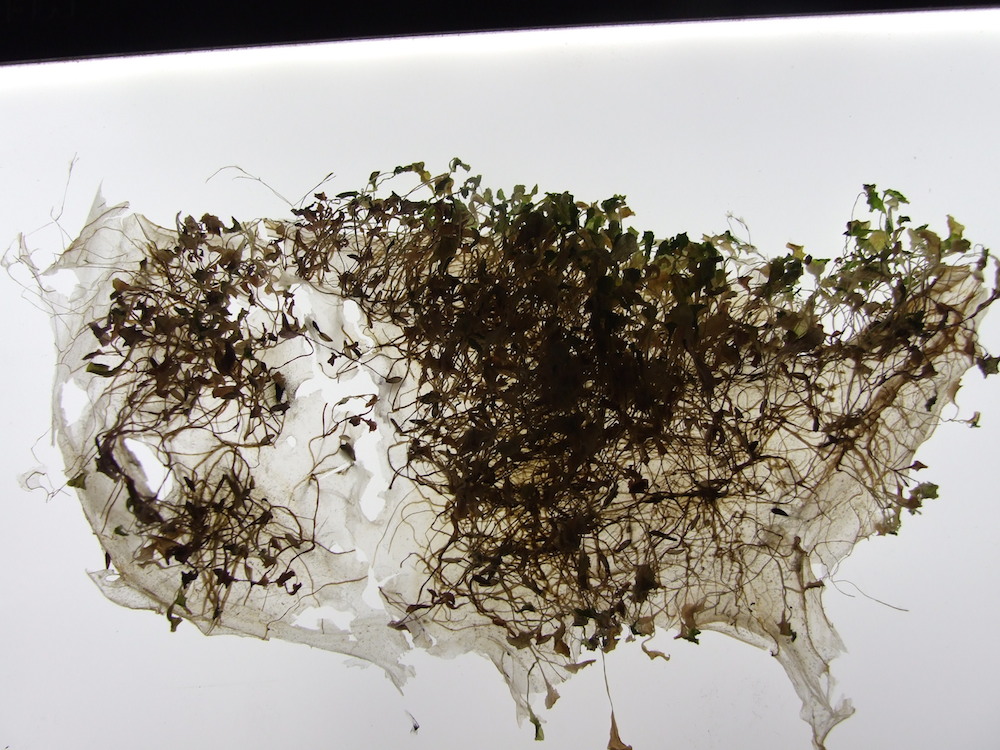}}
\subfigure[]{\includegraphics[width=0.48\textwidth]{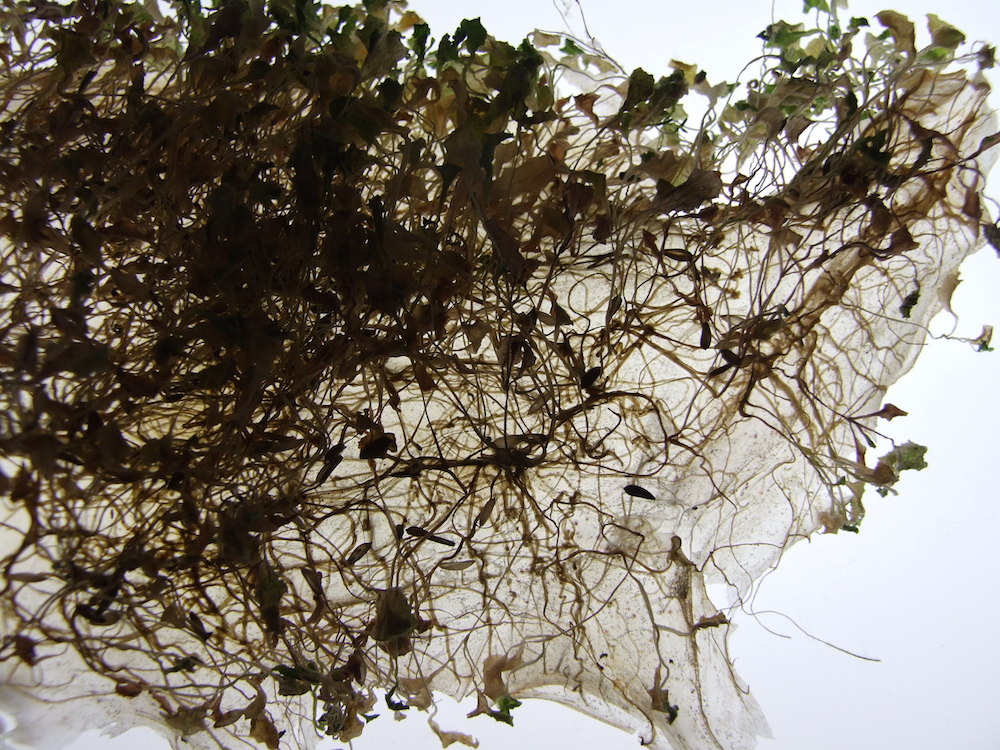}}
\caption{Plant roots on 3D templates of the USA. (a)~Maize seeds was placed at the location of the template corresponding to Minneapolis (the location was chosen for no apparent reason). Photo is made c. 7-10 days after inoculation.
(bcd)~Lettuce seeds were scatted in the USA template. (b)~Dried template with lettuce seedlings.
(c)~Agar film removed from the template. (d)~Zoomed part of the agar film representing eastern part of USA.}
\label{lettuceUSA}
\end{figure}

Collision-free path finding of plant roots was tested in further experiments in the Unconventional Computing Centre (UWE, Bristol). We have 3D printed with nylon templates of several countries, with elevation. Afterwards, we have been placing seeds either at bare templates or templates coated by 2\% phytogel. Templates were kept in containers with very high humidity for up to several weeks. The templates rested horizontally: there was no preferentially directions of root growth, neither directed by gravity or light nor by chemo-attractants. Therefore roots explored the space. A single maize seed placed at the position roughly corresponding to Minneapolis, produced several roots. The roots propagated in different from each other directions thus optimising space explored (Fig.~\ref{lettuceUSA}a). It seems that while navigating tips of roots avoid elevations: sometimes they go around localised elevation, sometimes they are reflected from the extended elevation. In another series of experiments, we scattered lettuce seeds on the template coated by phytogel. While scattering the seeds we were aiming to roughly approximate density of USA population: regions with high population density received a higher number of the seeds (Fig.~\ref{lettuceUSA}b). We observed that roots tend to cluster in groups (Fig.~\ref{lettuceUSA}c). This is a well known phenomenon \cite{ciszak2012swarming} and groups of routes propagate along valleys or river beds, definitely avoiding any elevations (Fig.~\ref{lettuceUSA}d). See also video of the experiment where lettuce seedlings grow on the 3D template of Russia:  \url{https://drive.google.com/open?id=0BzPSgPF_2eyUYlNlWEVwenVoLUU}.

\subsection{Spanning trees}
\label{spanningtrees}

\begin{figure}[!tbp]
\centering
\subfigure[]{\includegraphics[width=0.49\textwidth]{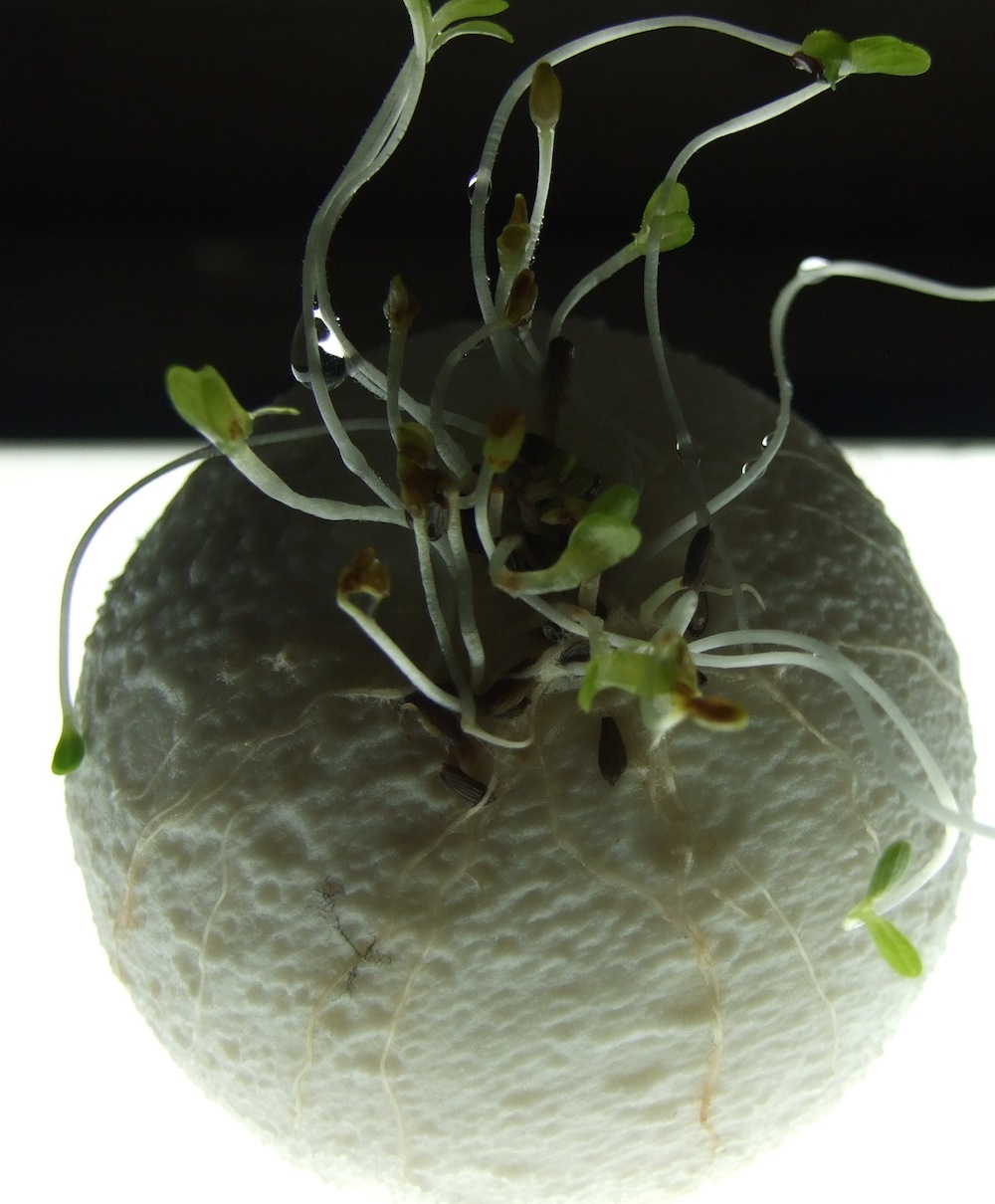}}
\subfigure[]{\includegraphics[width=0.49\textwidth]{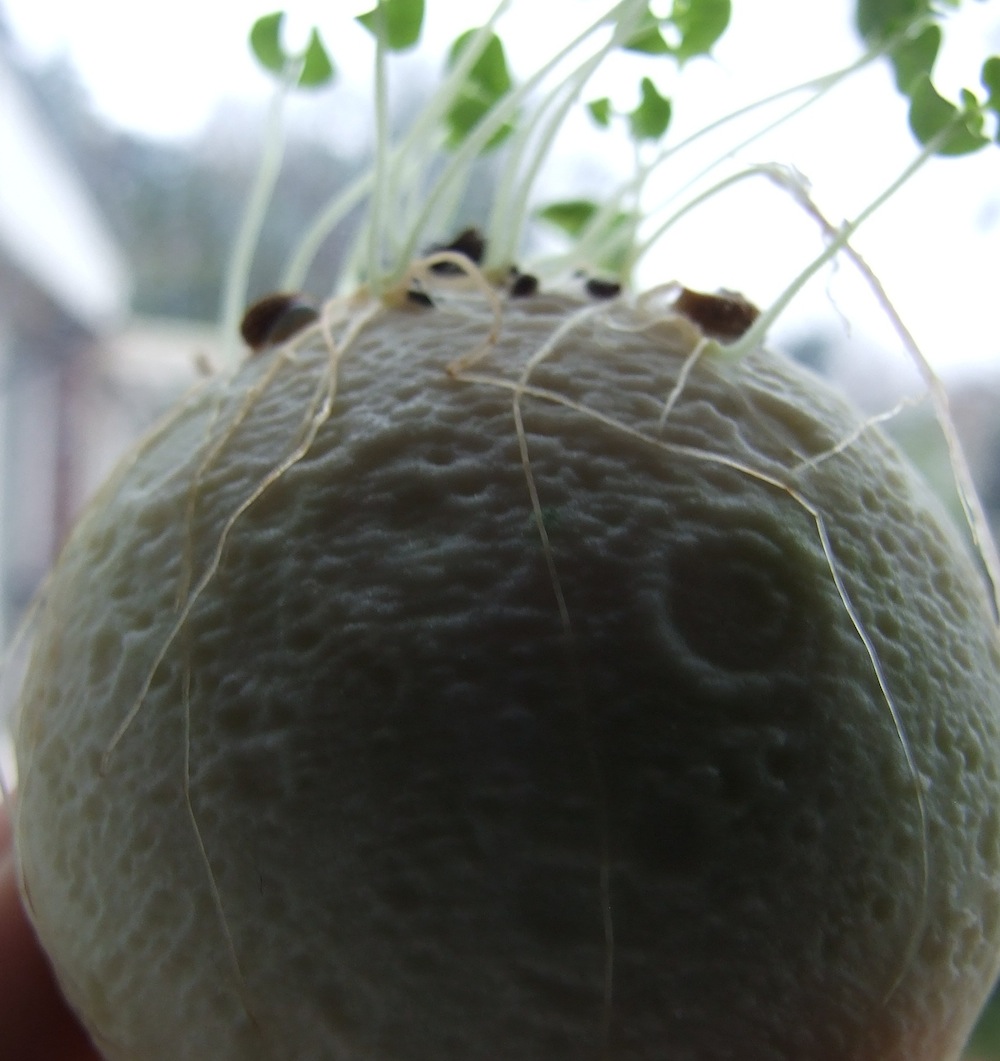}}
\caption{Plant roots explore surface of 3D template of Moon. 
(a)~Lettuce.
(b)~Basil.
}
\label{lettucemoon}
\end{figure}

Minimum spanning tree \index{minimum spanning tree} is a skeleton architecture of communication, transport and sensor networks. The Euclidean minimum spanning tree is a connected acyclic graph which has minimum possible sum of edges' lengths. To construct a spanning tree we can represent points of a data set with sources of nutrients and place a plant seed at one of the data points. The root will grow following gradients of the chemo-attractants, branch in the sites corresponding to the planar data set and, eventually, span the set with its physical body.
In 
\cite{adamatzky2014JBIS} we used live slime mould \emph{Physarum polycephalum} to imitate exploration of planets and to analyse potential scenarios of developing transport networks on Moon and Mars. In that experiments we have been inoculating the slime mould on 3D templates of the planets, at the positions matching sites of Apollo or Soyuz landings; then we allowed the slime mould to develop a network of protoplasmic tubes. We have repeated similar experiments but used plant seeds instead of slime mould (Fig.~\ref{lettucemoon}). We found that roots could be as good as the slime mould in imitating exploratory propagation, or scouting, in an unknown terrains. The basic traits include maximisation of geographical distance between neighbouring roots and avoidance of elevations.

\subsection{Crowd dynamics} 
\label{crowddynamics}

\begin{figure}[!tbp]
\centering
\subfigure[]{\includegraphics[width=0.58\textwidth]{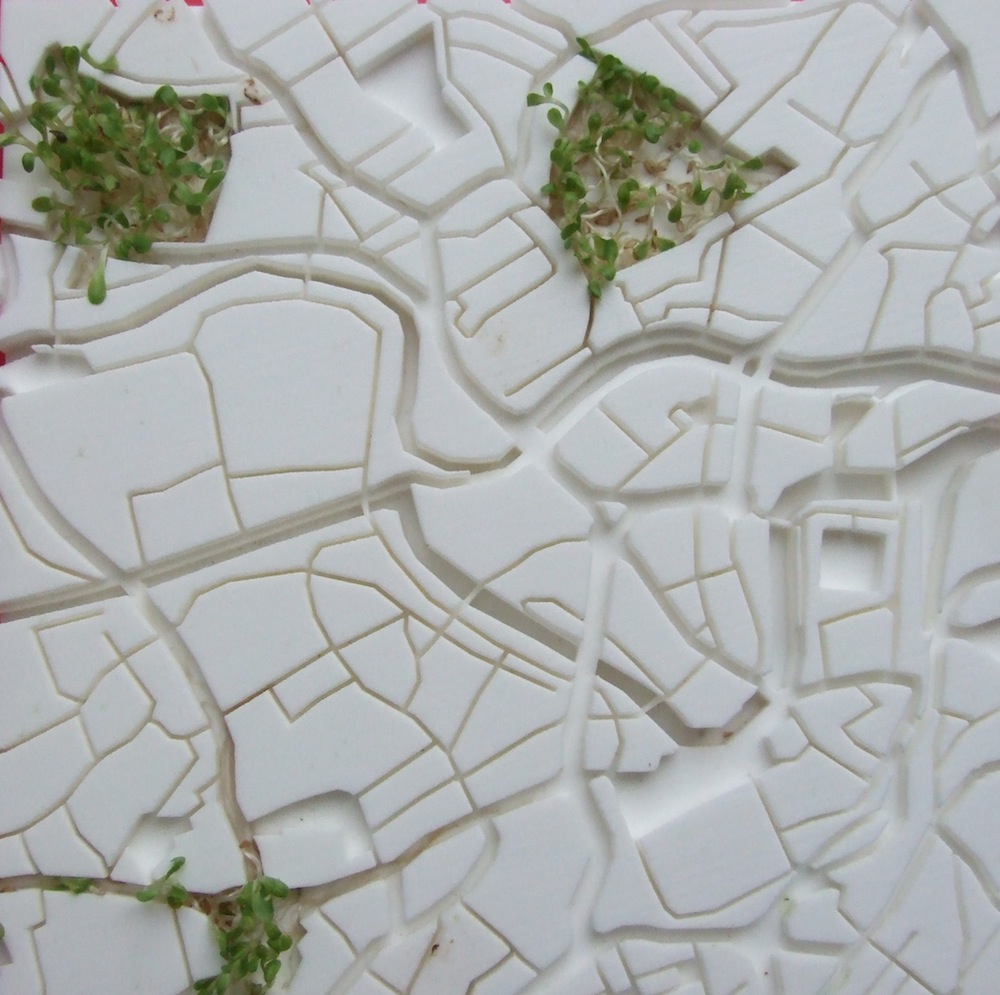}}
\subfigure[]{\includegraphics[width=0.58\textwidth, angle=90]{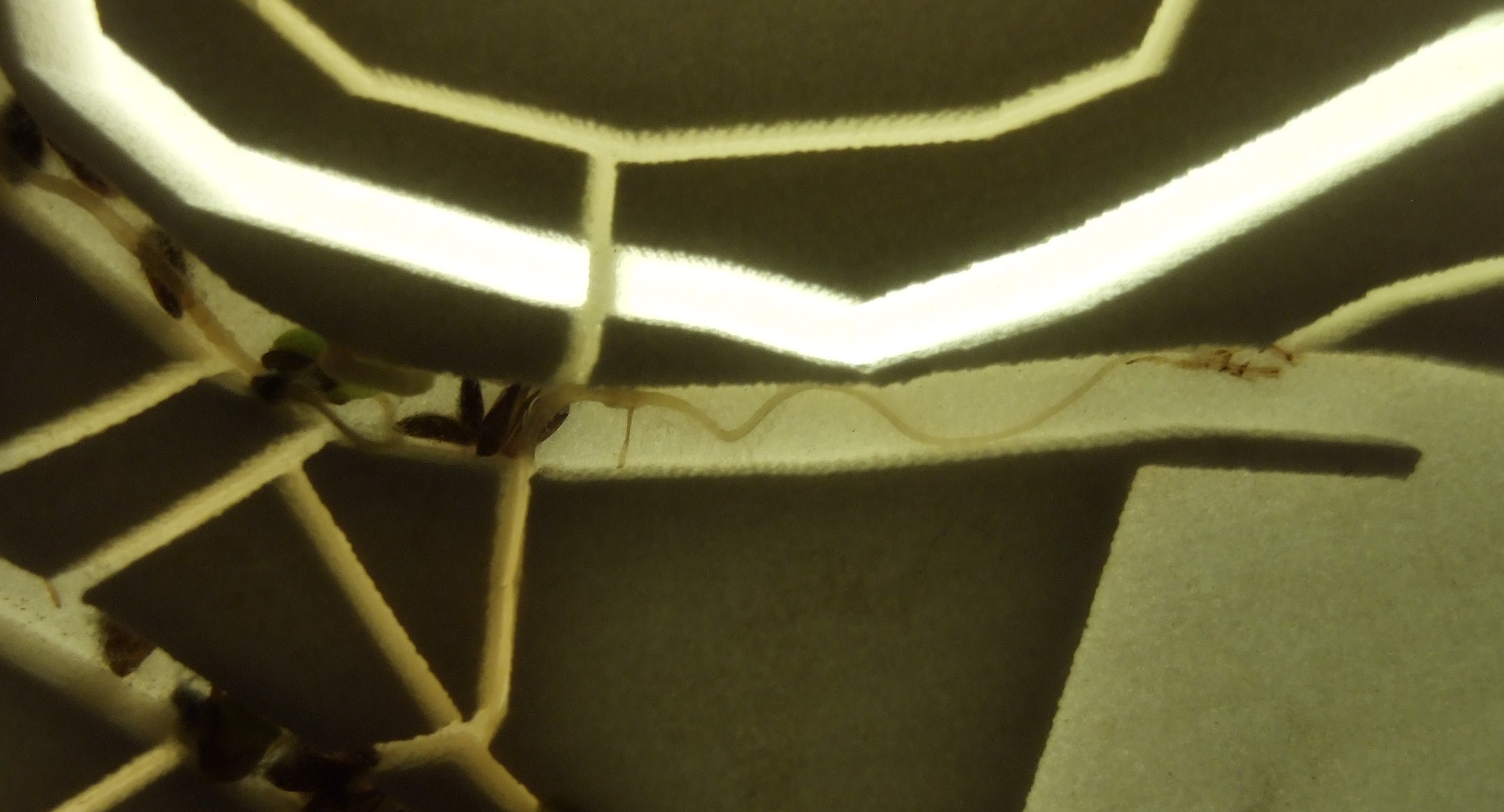}}
\subfigure[]{\includegraphics[width=0.48\textwidth]{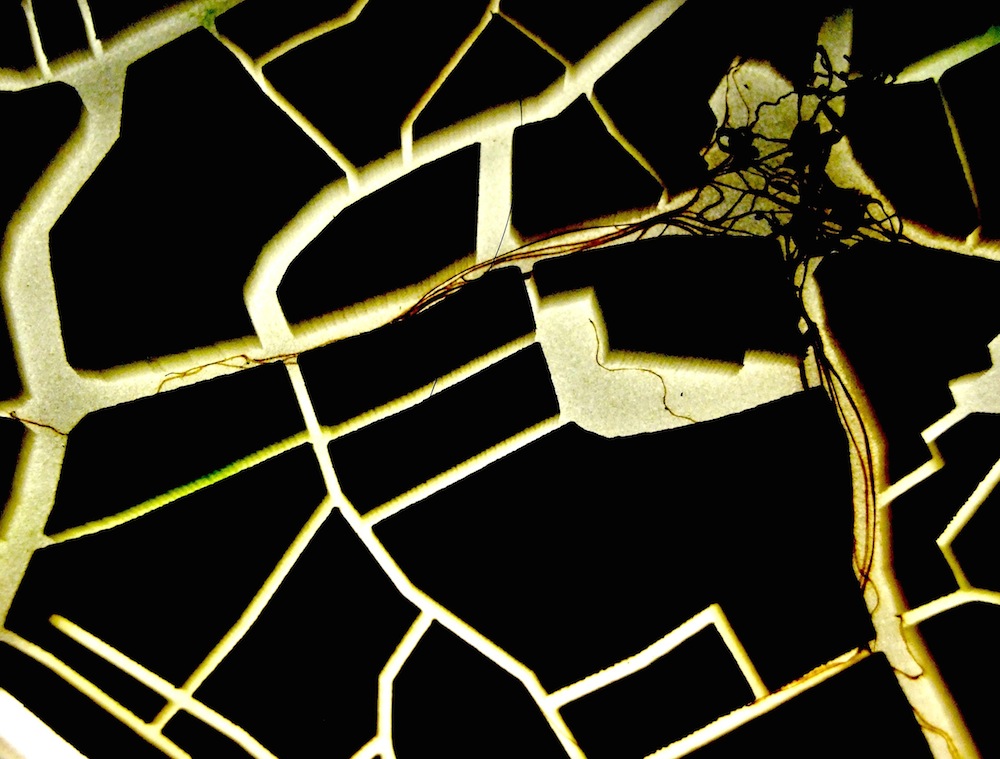}}
\subfigure[]{\includegraphics[width=0.48\textwidth]{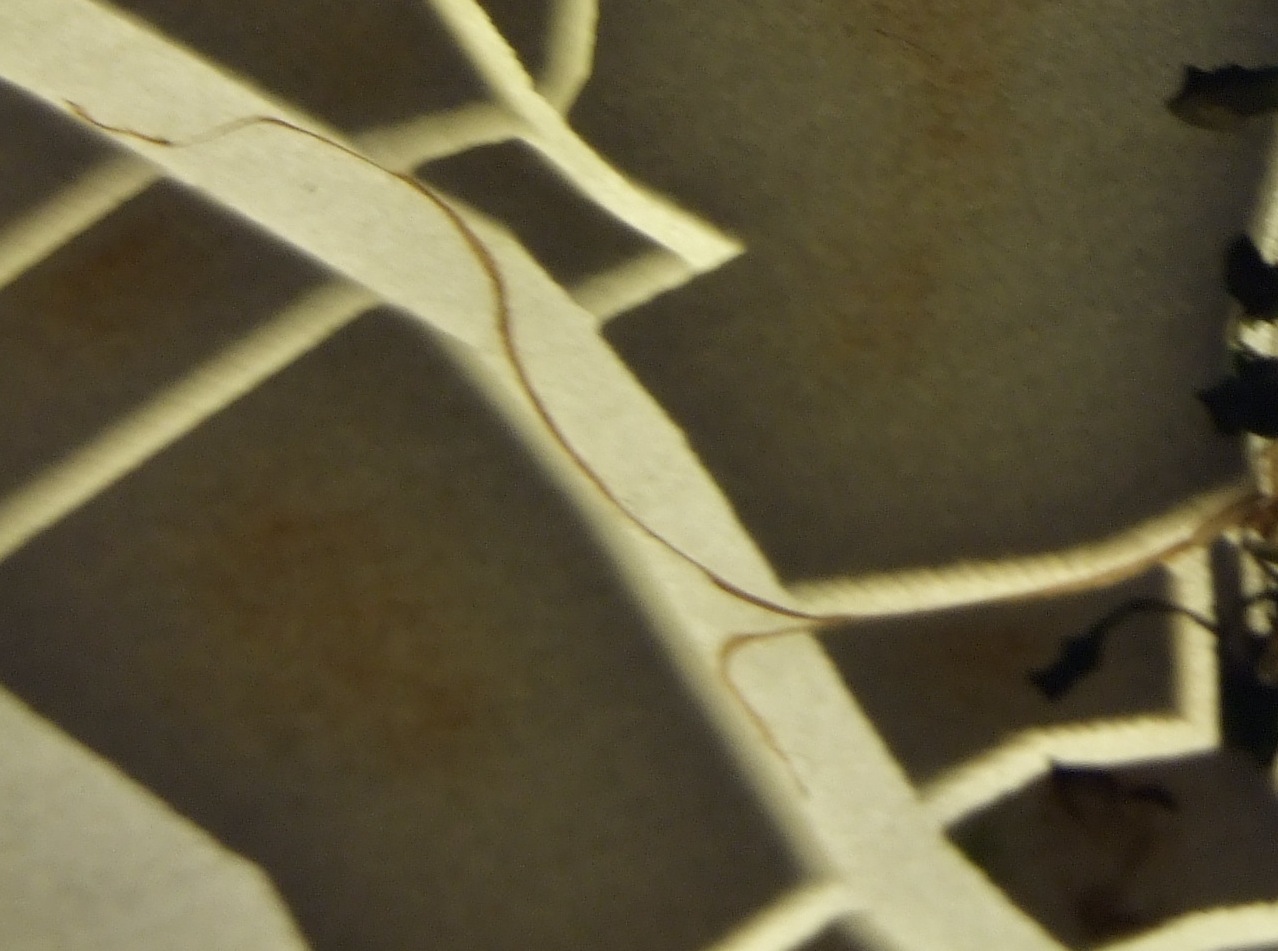}}
\caption{Imitation of crowd propagation with plant roots on a 3D template of Bristol, UK.
(a)~Seedlings are growing in three open spaces, including Temple Gardens and Queen Square.
(b)~Example of bouncing movement of root apex.
(c)~Root apexes form two swarms separated by an obstacle.
(d)~Two apexes repel each other and choose different directions of propagation.
}
\label{lettucebristol}
\end{figure}

Can we study crowd dynamics \index{crowd dynamics} using roots growing in a geometrically constrained environment? 
We printed a 3D nylon template of a part of Bristol city (UK), roughly a rectangular domain 
2 miles $\times$ 2 miles, centered at Temple Mead Train station. In this template, blocks of houses 
were represented by elevations and streets as indentations. To imitate crowds we placed seeds of 
lettuce in large open spaces, mainly gardens and squares (Fig.~\ref{lettucebristol}a). The templates were kept in a horizontal position in closed transparent containers with very high moisture contents. 
Morphology of growing roots were recorded in 7-14 days after start of experiments. We have made the following observations so far.

Root apexes prefer wider streets, they rarely (if ever) enter side streets, and narrowing branches of main streets (Fig.~\ref{lettucebristol}bcd). This may be explained by the fact that plants emit ultrasound \cite{perel2006ultrasound} and root apexes can sense ultrasound waves \cite{gagliano2012acoustic}. Chances are high that root apexes navigate in their constrained environment similarly to bats. Therefore entries to narrow streets are not detected.

In absence of attractants and repellents root apexes propagate ballistically: after entering a room a root grows along its original `velocity' vector until it collides with an obstacle. The apex reflects on collision. This is well illustrated in Fig.~\ref{lettucebristol}b. 
 
Root apexes swarm when propagating along wide streets (Fig.~\ref{lettucebristol}c), their growth is coherent, they often propagate in parallel, forming arrays of roots. Roots swarming is a known fact \cite{ciszak2012swarming} yet still a valuable observation in the context of imitating crowds.
 
Rays of apexes often dissipate on entering the wider space as illustrated in Fig.~\ref{lettucebristol}d. As shown, two roots propagate westward along a narrow street. Then they enter main street and collide with its west side. On `impact', one root deflects north-west another south-west.

\subsection{Voronoi diagram} 
\label{Voronoidiagram}

 \begin{figure}[!tbp]
\centering
\subfigure[]{\includegraphics[width=0.32\textwidth]{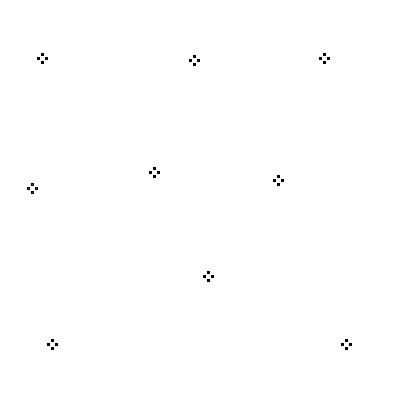}}
\subfigure[]{\includegraphics[width=0.32\textwidth]{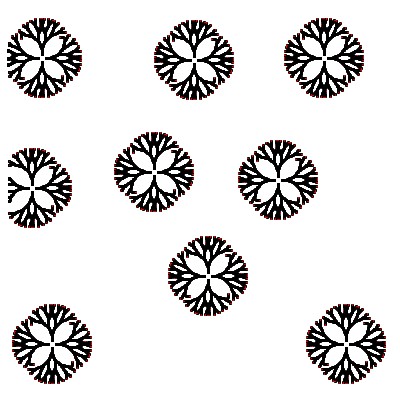}}
\subfigure[]{\includegraphics[width=0.32\textwidth]{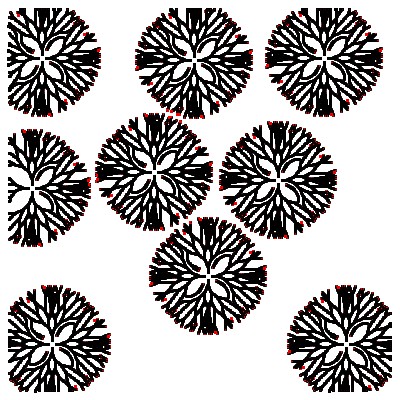}}
\subfigure[]{\includegraphics[width=0.32\textwidth]{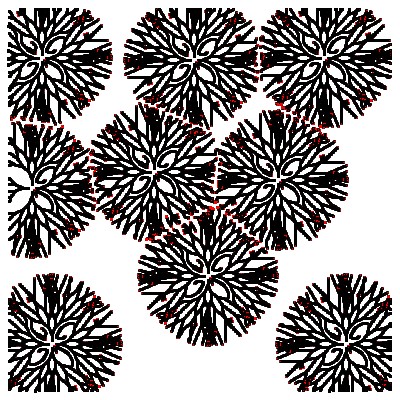}}
\subfigure[]{\includegraphics[width=0.32\textwidth]{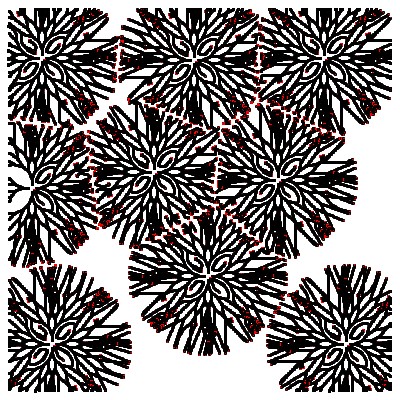}}
\subfigure[]{\includegraphics[width=0.32\textwidth]{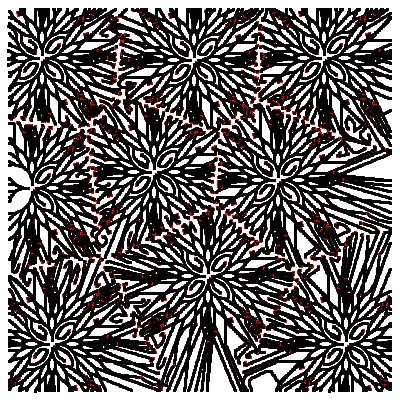}}
\caption{Approximation of Voronoi diagram with growing and branching roots. See details of computer model in \cite{adamatzky2011pluscomputing}. Snapshots are taken at (a)~160, (b)~216, (c)~234, (d)~246, (e)~256 and (f)~331 step of simulation. 
}
\label{voronoi}
\end{figure}

Let $P$ be a non-empty finite set of planar points. A planar Voronoi diagram \index{Voronoi diagram} of the set $P$ is a partition of the plane into such regions that, for any element of $P$, a region corresponding to a unique point $p$ contains all those points of the plane which are closer to $p$ than to any other node of $P$. The planar diagram is combinatorially equivalent to the lower envelope of unit paraboloids centred at points of $P$. Approximation of the following Voronoi diagrams --- planar diagram of point set, generalised of arbitrary geometrical shapes, Bregman diagrams on anisotropic and inhomogeneous spaces, multiplicative and furthest point diagrams --- could be produced when implemented in experimental laboratory conditions with roots.

The Voronoi diagram can be approximated by growing roots similarly to the approximation of the diagram with slime mould~\cite{adamatzky2010physarum} or precipitating reaction-diffusion fronts~\cite{adamatzky2005reaction}. We represent every point of a data set $P$ by a seed. The branching root system grows omnidirectionally. When wave-fronts of growing root systems, originated from different seeds, approach each other they stop further propagation. Thus loci of space not covered by roots represent segments of the Voronoi cells. The approach can be illustrated using a model of growing and branching pattern, developed by us originally to imitate computation with liquid crystal fingers~\cite{adamatzky2011pluscomputing}. Seeds are places in data points (Fig.~\ref{voronoi}a), root growth fronts propagate (Fig.~\ref{voronoi}b--c), collide with each other (Fig.~\ref{voronoi}d,e). The Voronoi diagram is approximated when the system becomes stationary and no more growth occurs (Fig.~\ref{voronoi}f).

\subsection{Planar hulls} 
\label{planarhulls} 
 
Computing a polygon defining a set of planar points is a classical problem of computational geometry. $\alpha$-hull \index{$\alpha$-hull} of a planar set $P$ is an intersection of the complement of all closed discs of radius 1/$\alpha$ that includes no points of $P$. $\alpha$-shape is a convex hull when $\alpha \rightarrow \infty$. We can represent planar points $P$ with sources of long-distance attractants and short-distance repellents and place a root outside the data set. The roots propagate towards the data and envelop the data set with their physical bodies. We can represent value of $\alpha$ by attractants/repellents with various diffusion constants and thus enable roots to calculate a wide range the shapes including concave and convex hulls. This approach worked well in experiments on approximation of concave hull with slime mould \cite{adamatzky2012slimehull}.

\subsection{Subdivision of concave polygons} 
\label{concavepolygon}
 
\begin{figure}[!tbp]
\centering
\subfigure[]{\includegraphics[width=0.24\textwidth]{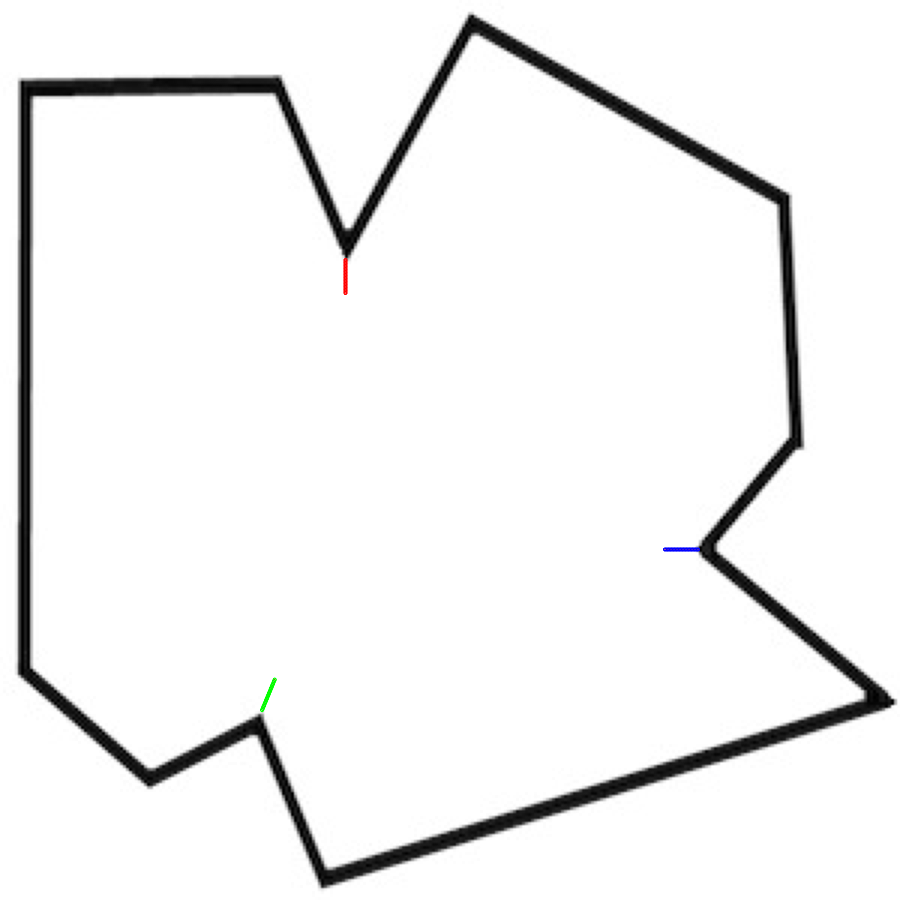}}
\subfigure[]{\includegraphics[width=0.24\textwidth]{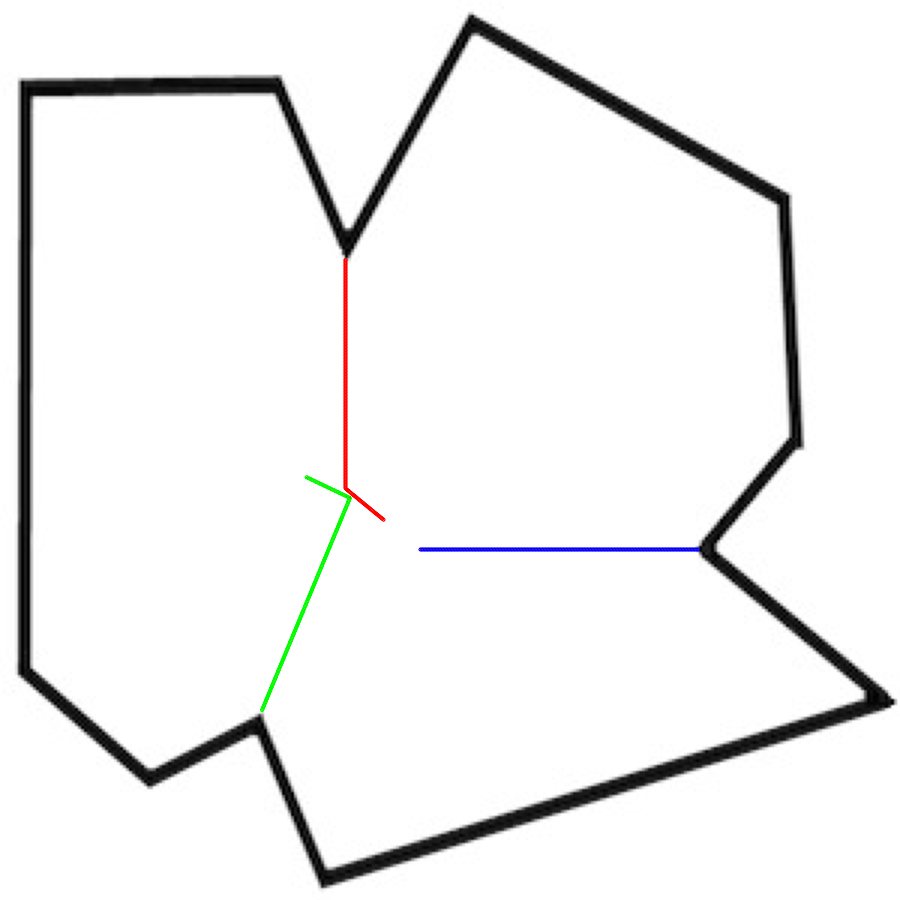}}
\subfigure[]{\includegraphics[width=0.24\textwidth]{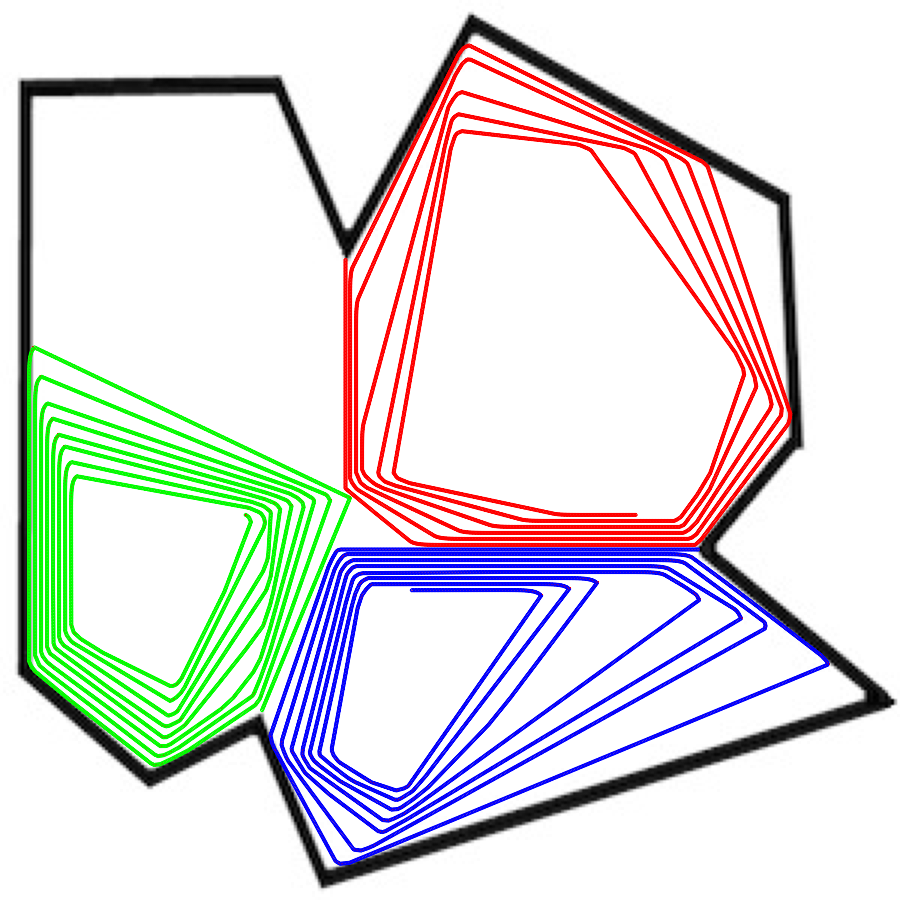}}
\subfigure[]{\includegraphics[width=0.24\textwidth]{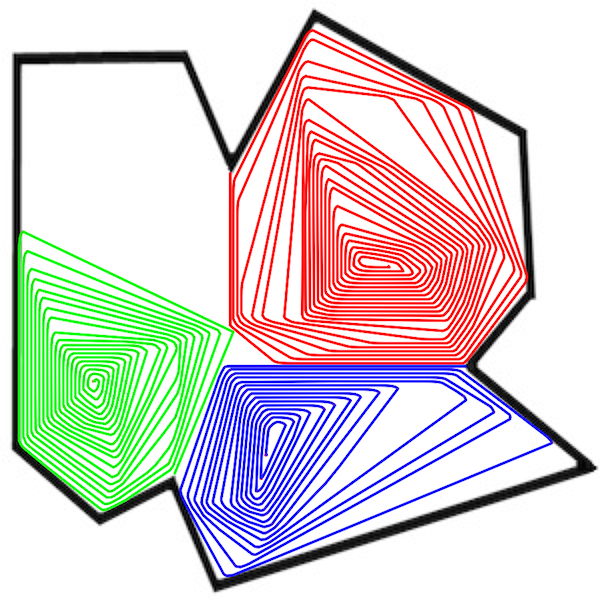}}
\caption{Illustration on how a polygon can be subdivided by plant roots. 
The original model refers to liquid crystal fingers~\cite{adamatzky2011pluscomputing} but mechanisms of interaction could be the same. Snapshots are taken at different stages of simulation, see details in ~\cite{adamatzky2011pluscomputing}}
\label{subdivision}
\end{figure} 
 
A concave polygon \index{concave polygon} is a shape comprised of straight lines with at least one indentation, or angle pointing inward. The problem is to subdivide the given concave shape into convex shapes. The problem can be solved with growing roots as follows. Roots are initiated at the singular points of indentations (of the data polygon) and the roots propagation vectors are co-aligned with medians of the corresponding inward angles. Given a concave polygon, every indentation initiates one propagating root apex. By applying an external electro-magnetic field we can make root apexes turning only left (relatively to their vector of propagation). By following this ``turn-left-if-there-is-no-place-to-go'' routine and also competing for the available space with each other, the roots fill $n-1$ convex domains. At least one convex domain will remain unfilled. See an example of subdivision of a concave polygon in Fig.~\ref{subdivision}.

\subsection{Logical gates from plant roots}
\label{logicalgatesplant}

A  collision-based computation, \index{collision-based computation} emerged from Fredkin-Toffoli conservative logic~\cite{fredkin2002conservative}, employs mobile compact finite patterns, which implement computation while interacting with each other~\cite{adamatzky2002collision}. Information values (e.g. truth values of logical variables) are given by either absence or presence of the localisations or other parameters of the localisations. The localisations travel in space and perform computation when they collide with each other.  Almost any part of the medium space can be used as a wire. The localisations undergo transformations, they change velocities, form bound states and annihilate or fuse when they interact with other localisations. Information values of localisations are transformed as a result of collision and thus a computation is implemented. 
\index{logical gate}
In \cite{adamatzky2016plant} we proposed theoretical constructs of logical gates implemented with plant roots as morphological computing asynchronous devices. 
Values of Boolean variables are represented by plant roots. A presence of a plant root at a given site symbolises the logical {\sc True}, an absence the logical {\sc False}. Logical functions are calculated via interaction between roots. Two types of two-inputs-two-outputs gates are proposed~\cite{adamatzky2016plant}: a gate $\langle x, y \rangle \rightarrow \langle xy, x+y \rangle$  where root apexes are guided by gravity and a gate $\langle x, y \rangle \rightarrow \langle \overline{x}y, x \rangle$ where root apexes are guided by humidity. Let us show how a logical gate based on attraction of roots can be implemented.

Root apexes are attracted to humidity~\cite{baluvska2004root} and a range of chemical compounds~\cite{schlicht2013indole, xu2013improved, graham1991flavonoid, bais2006role, steinkellner2007flavonoids, yokawa2014binary}. A root apexes grow towards the domain with highest concentration of attractants. The root minimises energy during its growth: it does not change its velocity vector if environmental conditions stay the same. This is a distant analog of inertia.

\begin{figure}[!tbp] 
\centering
\includegraphics[width=\textwidth]{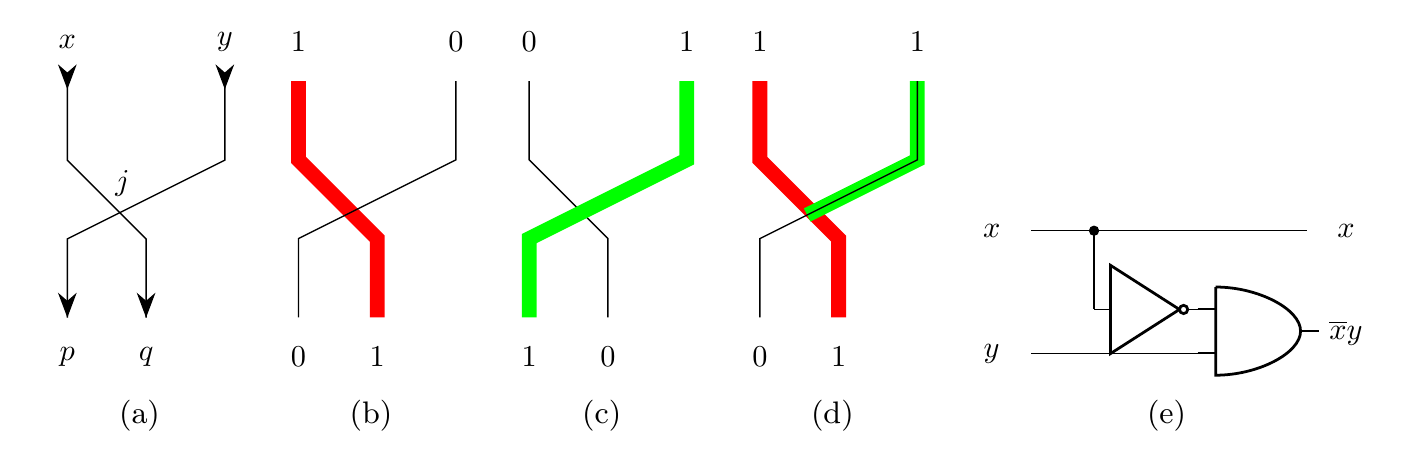}
\caption{(a)~Scheme of humidity gate with two inputs $x$ and $y$ and two outputs $p$ send $q$: $p=\overline{x}y$ and $q=x$.
(b)~$x=1$ and $y=0$. 
(c)~$x=0$ and $y=1$. 
(d)~$x=1$ and $y=1$.
(e)~Equivalent logic scheme. From \cite{adamatzky2016plant}.}
\label{humidity}
\end{figure}

\begin{figure}[!tbp] 
\centering
\includegraphics[width=\textwidth]{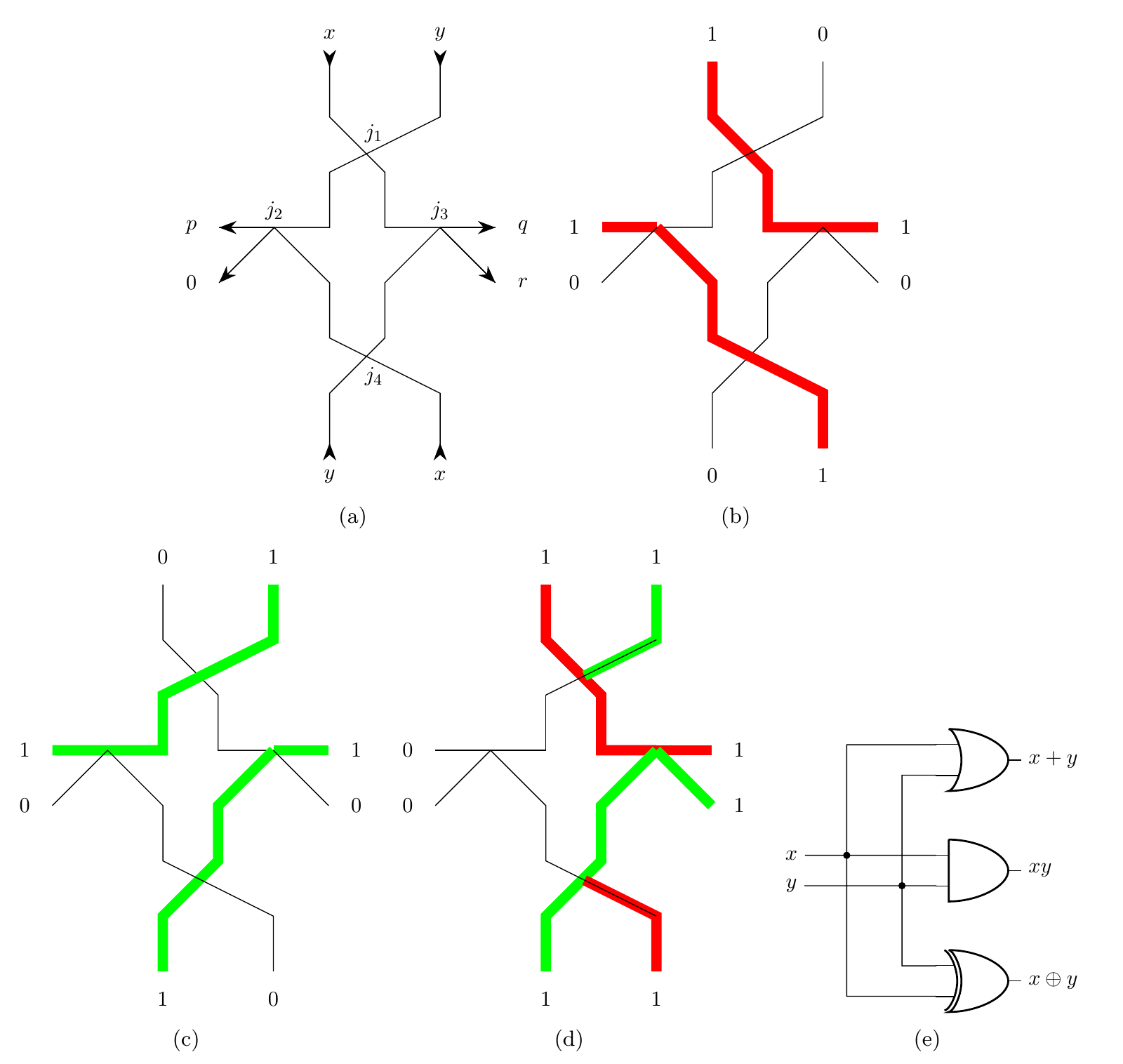}
\caption{A half-adder made of two humidity gates. 
(a)~Scheme of the circuit, $p= x \oplus y$, $q=x+y$, $r=xy$.
(b)~$x=1$ and $y=0$. 
(c)~$x=0$ and $y=1$. 
(d)~$x=1$ and $y=1$.
(e)~Equivalent logic gates design.From \cite{adamatzky2016plant}.}
\label{humidityadder1}
\end{figure} 

Assume attractants are applied at the exits of channels $p$ and $q$ (Fig.~\ref{humidity}a). When an apex of the root, growing along channel $x$, reaches a junction between channels, the apex continues (due to energy minimisation) its growth into the channel $q$ if this channel is not occupied by other root (Fig.~\ref{humidity}b). A root in input channel $y$ grows through the junction into the output channel $p$ (Fig.~\ref{humidity}c). 

The gate Fig.~\ref{humidity}a has such a geometry that a path along channel $x$ to junction $j$ is shorter than a path along channel $y$ to the junction $j$. Therefore, a root growing in channel $x$ propagates through the junction into channel $q$ before root starting in channel $y$ reaches the junction. When both roots are initiated in the input channels the $x$-root appears in the output $q$ but the $y$-root is blocked by the $x$-root from propagating into the channel $p$: no signal appears at the output $p$ (Fig.~\ref{humidity}d). 
This gate realises functions $p=\overline{x}y$ and $q=x$ (Fig.~\ref{humidity}e).  If $y$ is always 1 the gate produces a signal and its negation at the same time.

Two attraction gates Fig.~\ref{humidity}a can be cascaded into a circuit to implement a one-bit half-adder, \index{one-bit half-adder} with additional output, as shown in Fig.~\ref{humidityadder1}a. We assume the planar gate is lying flat and sources of attractants are provided near exits of output channels $p$, $q$ and $r$. The half-adder is realised on inputs $p=x \oplus y$ (sum) and $r=xy$ (carry); the circuit has also a  `bonus' output $q=x+y$ (Fig.~\ref{humidityadder1}e). The circuits work as follows:
\begin{itemize}
\item Inputs $x=1$ and $y=0$: Two roots are initiated in channels marked $x$ in Fig.~\ref{humidityadder1}a; 
one root propagates to junction $j_1$ to junction $j_3$ and exits in channel $q$; 
another root propagates to junction $j_4$ to junction $j_2$ and into channel $p$ (Fig.~\ref{humidityadder1}b). 
\item Inputs $x=0$ and $y=1$: Two roots are initiated in channels marked $y$ in Fig.~\ref{humidityadder1}a;
one root propagates to junction $j_1$ then to junction $j_2$ and exits at channel $p$;
another root propagates to junction $j_4$ then to junction $j_3$ then into channel $q$ (Fig.~\ref{humidityadder1}c).
\item Inputs $x=1$ and $y=1$:  Roots are initiated in all four input channels. The root initiated in the northern channel $x$ propagates towards exit $q$. This root blocks propagation of the root initiated in the southern channel $y$, therefore the root from the southern channel $x$ exits the circuit via the channel $r$. The root growing 
in the northern channel $x$ blocks propagation of the root initiated in the norther channel $y$, therefore no roots appear in the output $p$ (Fig.~\ref{humidityadder1}d).
\end{itemize}

\section{Plant electronics}
\label{plantelectronics}

In \index{plant electronics} living electronic processors parts of a plant are functionalized via coating with polymers and loading with nano-particles, thus local modulations of the plant's’ electrical properties is achieved. A whole plant is transformed into an electronic circuit, where functionalized parts of  the plant are basic electronic elements and 'raw' parts are conductors.

\subsection{Plant wire}
\label{plantwire}

In \cite{adamatzky2014towards} we have exhibited that plants can function as wires, \index{plant wire} albeit slightly noisy ones. Namely, in laboratory experiments with lettuce seedlings we found that the seedlings implement a linear transfer function of input potential to output potential. Roughly an output potential is 1.5--2~V less than an input potential, thus e.g. by applying 12~V potential, we get 10~V output potential. Resistance of 3-4 day lettuce seedling is about 3~M$\Omega$ on average. This is much higher than resistance of conventional conductors yet relatively low compared to the resistance of other living creatures~\cite{geddes1967specific}. In our experiments \cite{adamatzky2014towards} we measured resistance by bridging two aluminium electrodes with a seedling. If we did insert Ag/AgCl needle electrodes inside the seedling, we would expect to record much lower resistance, as has been shown in \cite{mancuso1999seasonal}.

\begin{figure}[!tbp]
\centering
\includegraphics[width=0.8\textwidth]{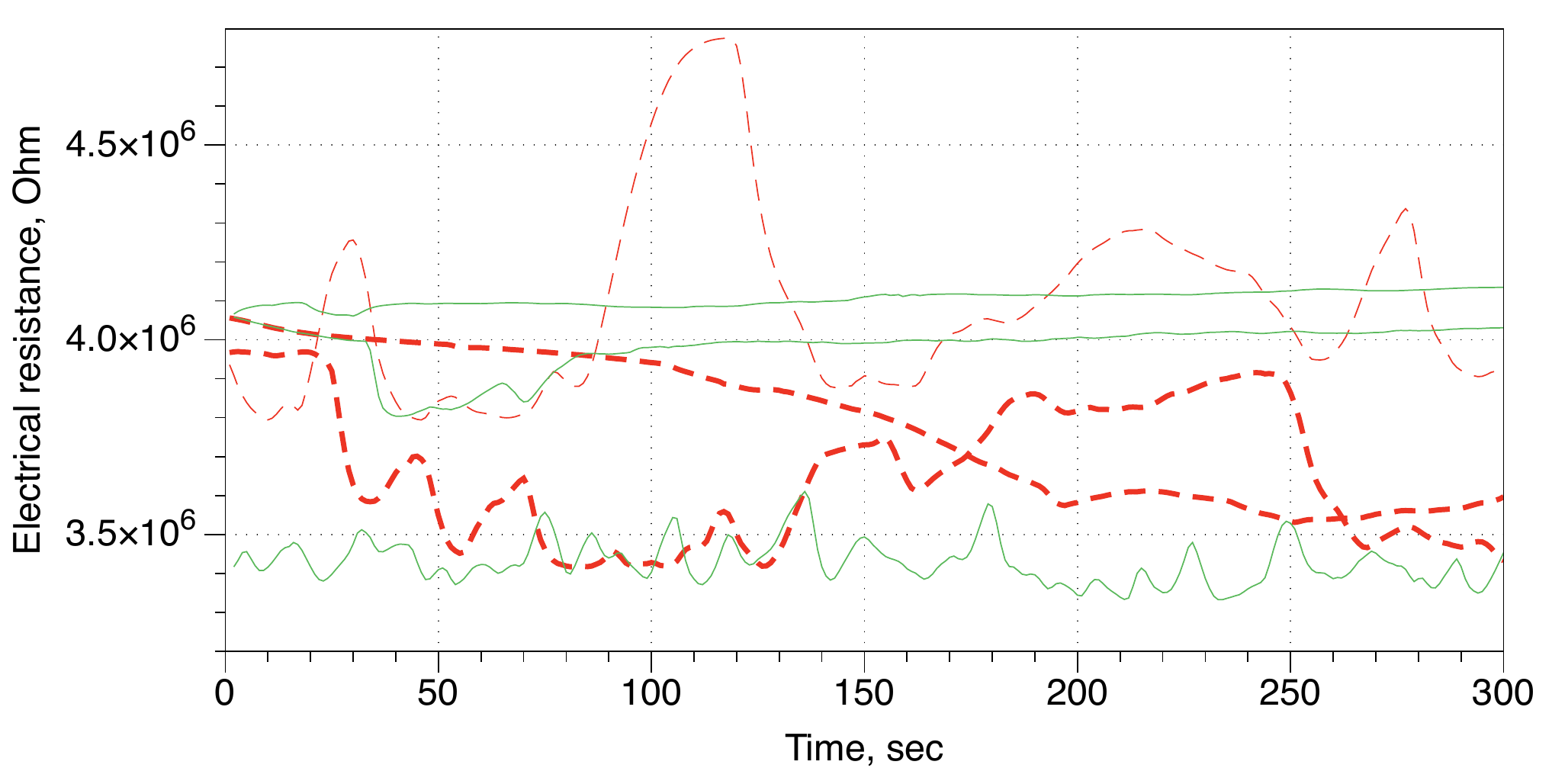}
\caption{Effect of chloroform on lettuce electrical resistance. Solid green lines show the resistance of an intact lettuce seedling and dashed red lines the resistance of a seedling in presence of chloroform.}
\label{chlorophorm}
\end{figure}

Resistance of plant wires can be affected by temperature~\cite{mancuso1999seasonal}, illumination and chemical substances. For example, when a lettuce seedling is exposed to vapour of chloroform (one 1~$\mu$L in a standard 90~mm Petri dish) the seedling exhibits high amplitude irregular oscillations of its resistance (Fig.~\ref{chlorophorm}).

\subsection{Functionalizing plants}
\label{functionalizingplants}

Electrical properties of plants can be changed by synthesis of inorganic materials by plants, coating roots with metal nano-particles and conductive polymers, growth of alloy networks. Here we overview general principles, some of them might not be applicable for plants, subject to further studies.

Many organisms, both unicellular and multicellular, produce inorganic materials either intra- or extra-cellular, e.g. include magneto-tactic bacteria (which synthesize magnetite nano-particles), diatoms (which synthesize siliceous materials) and S-layer bacteria (which produce gypsum and calcium carbonate layers). Biomimetic inorganic materials have been recently achieved by morpho-synthesis of biological templates such as viruses, bacteria, diatoms, biopolymers, eggshells, sea urchins, spider silks, insects, wood and leaves. We can employ recent results in biomorphic mineralization to produce reusable components of plants with conductive and magnetic substrates.

To coat roots with gold nano-particles (intake, transport and excretion) we can proceed as follows. Gold nano-particles (20-500 nm) pure or coated with bio-affine substances can be saturated in the feeding substrate of roots and/or applied as a liquid directly to growing roots. The gold particles will be in-taken by the roots, transported along the roots, distributed in the root network and eventually excreted onto the outer cell wall of root cells. When roots cease to function, the gold coating stays in place, providing passive conducting pathway. 

Cellular synthesizes can be implemented via in plant growth of (semi-)conductive crystals from metal ions present in substrate. The growing roots will recover silver particle extracellularly from solutions containing Ag$^+$ ions, e.g. by saturating roots in 1~mM soluble silver in the log phase of growth. We can expose roots to aqueous AuCl$^{4-}$ ions; the exposure will result in reduction of the metal ions and formation of gold nano-particles of around 20 nm.

To growth alloy networks of plants we can use two approaches. 
First, co-growing of single metal networks. One root network is coated gold, and then another root network is grown on top of it and coated with silver. 
Another approach could be to synthesise nano-materials with programmed morphology. Some preliminary studies with biological substrates, were successful. Alloy nano-particles exhibit unique electronic, optical, and catalytic properties that are different from those of the corresponding individual metal particles. We can expose plants to equi-molar solutions of HAuCl$_4$ and AgNO$_3$ so that the formation of highly stable Au-Ag alloy nano-particles of varying mole fractions could be achieved.

Polyaniline (PANI) \index{polyaniline} is a conducting polymer that can reach very high level of the conductivity (about 50-100 S/cm in our experiments), several forms of PANI can be deposited using layer-by-layer technique. Next layer is electro-statically attracted and, again, its thickness growth is blocked when the previous charge is compensated and new charge prevents further absorption due to the electrostatic repulsion. We can coat surface of roots with PANI, with a control of the thickness of about 1~nm. The approach has been successfully tested on several biological objects, notable slime mould~\cite{cifarelli2015bio, dimonte2016physarum, battistoni2017organic}. When implementing passive electrical components from plants we can take into account the geometry and morphology of the roots network coated with PANI. The electrical conductivity of the signal pathway will depend only on the length and level of branching of the connection according to the Ohm's law. Of course, we will be able to set the basic level of the conductivity varying the thickness of the deposited layer. With regards to active electrical behaviour, conductivity state of PANI dependents on its redox state. Oxidized state is conducting and reduced state is insulating. The difference in conductivity is about 8 orders of magnitude. Thus, the conductivity of the individual zone of PANI will depend also on its actual potential (the use of plant roots implies the presence of the electrolyte, that will act as a medium for the redox reactions).The conductivity map will depend not only on the morphology, but also on the potential distribution map, that is connected to the previous function of the network. An attempt could be made to produce a simple bipolar junction transistor, with possible extension to an operational amplifier.

Another important feature of PANI layers, that can be useful for the plant computers, is its capability to vary the color according to the conductivity state~\cite{battistoni2016spectrophotometric}. This property will allow to register the conductivity map of the whole formed network, while usually we have the possibility to measure only between fixed points where electrodes are attached. If we are thinking about the system with learning properties~\cite{demin2015hardware}, it can be not enough – we must register the variation of the connections between all elements of the network. In the case of a double-layer perceptron, for example, it required a realization of rather complicated external electronic circuit, providing the temporal detachment of individual elements from the network for the registering of its conductivity~\cite{emelyanov2016first}. Instead, the application of the spectroscopic technique allows monitoring of the conductivity state of all elements of the network in a real time for rather large area (up to half a meter), what will simplify, for example, the application of back propagation learning algorithms.

\subsection{Case study. Modifying lettuce with nanomaterials}
\label{casestudy}

\begin{figure}[!tbp]
\centering
\subfigure[]{\includegraphics[width=0.49\textwidth]{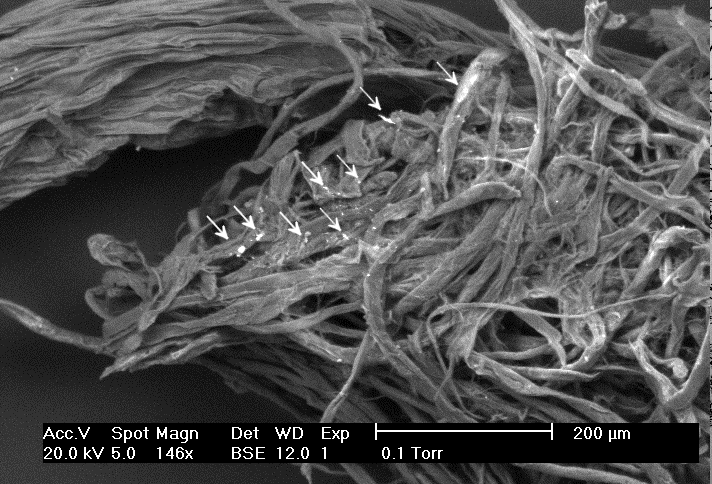}}
\subfigure[]{\includegraphics[width=0.49\textwidth]{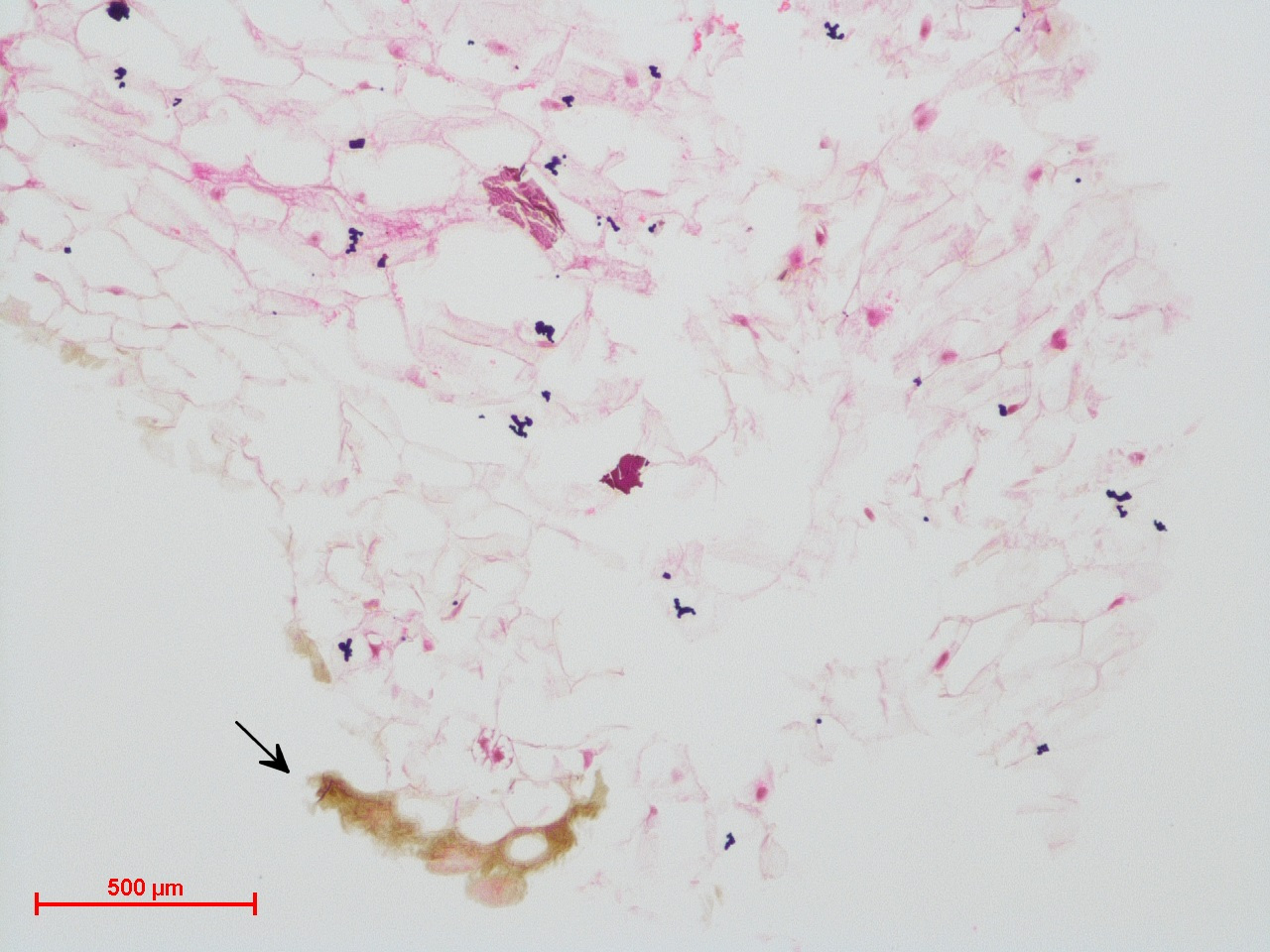}}
\caption{Lettuce functionalised with nanomaterials. 
(a)~Scanning electron micrograph and EDX spectrum of lettuce seedlings treated with aluminium oxide: low magnification of a snapped lettuce seedling stem showing bright regions. 
(b)~Light micrograph of 4~$\mu$m sections of lettuce seedlings treated with graphene in transverse orientation, haemotoxylin and eosin staining.
From \cite{gizzie2016hybridising}.
}
\label{fig:SEM}
\end{figure}

In \cite{gizzie2016hybridising} we hybridised lettuce seedlings with a variety of metallic and non-metallic nanomaterials; carbon nanotubes, graphene oxide, aluminium oxide and calcium phosphate. \index{carbon nanotubes} \index{graphene oxide} 
\index{aluminium oxide} \index{calcium phosphate} Toxic effects and the following electrical properties were monitored; mean potential, resistance and capacitance. Macroscopic observations revealed only slight deleterious health effects after administration with one variety of particle, aluminium oxide. 

\begin{figure}[!tbp]
\centering
\includegraphics[width=0.6\textwidth]{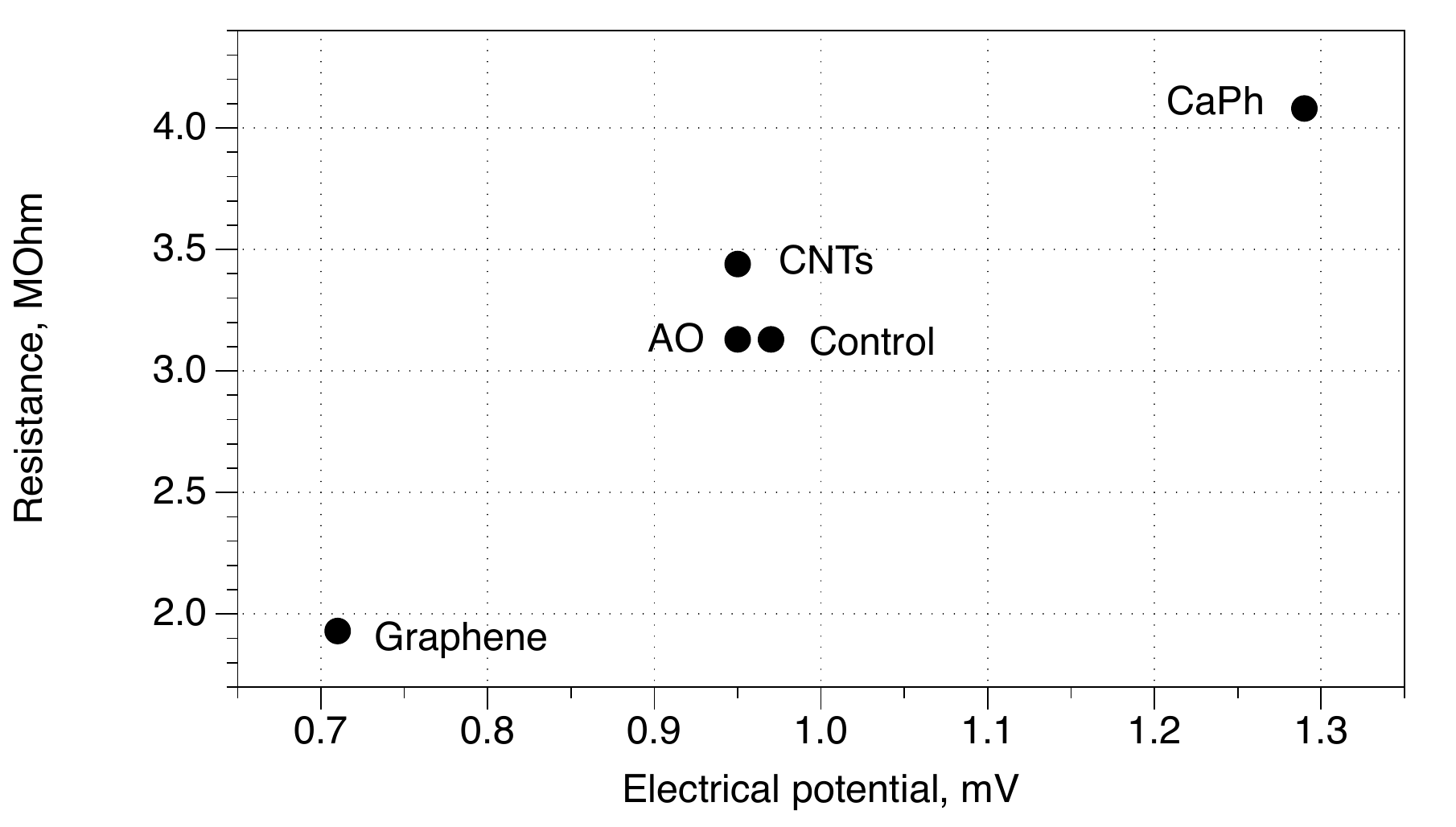}
\caption{Modifying lettuce electrical properties: resistance versus electrical potential plot. Nanomaterials used are graphene, carnon nanotybes (CNTs), calcium phosphate (CaPh) and aluminimum oxide (AO). From \cite{gizzie2016hybridising}.}
\label{LettuceModes}
\end{figure}

\begin{table}[!tbp]
\caption{Effects of selected nanomaterials on lettuce seedlings. If intake of the nanomaterial increases the measured parameters we indicate $\uparrow$, if it decreases $\downarrow$ and if parameter is within 5\% of the control then the value is recorded as unchanged $0$.  If parameter is altered 25\% above or below the control the arrow signs are encircled. Nanomaterials used are graphene, carnon nanotybes (CNTs), calcium phosphate (CaPh) and aluminimum oxide (AO). From \cite{gizzie2016hybridising}.}
\centering
\begin{tabular}{lcccc}
\hline
Material        & Potential & Resistance & Capacitance  \\
\hline
Graphene    &  $\downarrow$ & \textcircled{$\downarrow$}  & $\uparrow$    \\ 
CNTs        &  0 & $\uparrow$ & 0   \\
CaPh       & \textcircled{$\uparrow$}  & $\uparrow$  & $\downarrow$ \\
AO          &  0 & 0 & $\uparrow$ \\
\hline
\end{tabular}
\label{table-sum}
\end{table}

Mean potential in calcium phosphate-hybridised seedlings showed a considerable increase when compared with the control, whereas those administered with graphene oxide showed a small decrease; there were no notable variations across the remaining treatments. Electrical resistance decreased substantially in graphene oxide-treated seedlings whereas slight increases were shown following calcium phosphate and carbon nanotubes applications. Capacitance showed no considerable variation across treated seedlings. These results demonstrate that use of some nanomaterials, specifically graphene oxide and calcium phosphate may be used towards biohybridisation purposes including the generation of living `wires'.

Graphene oxide and calcium phosphate were found to be, by a margin of at least 25\%, the strongest modulators of the natural electrical properties of lettuce seedlings in this study, as is summarised in Fig.~\ref{LettuceModes} and  Table~\ref{table-sum}; although statistical significance was not achieved between the control and the nanomaterials, statistical significance was achieved between the resistance changes of graphene oxide and calcium phosphate. 

We also provided evidence that nanomaterials are able to enter different histological layers of the plant, through demonstrating that graphene oxide and latex spheres become lodged in the epidermis whereas aluminium oxide and some latex spheres travel into the stem. With pores of plants being up to 8~nm in diameter, the size of the nanoparticles must be important when hybridising with lettuce seedlings. However, large nanomaterials may be able to embed and coat their surface, which provides extra benefits as any toxic effects caused by the nanomaterials would be reduced as there is less interference with proteins and intracellular mechanisms. Toxicity may also vary dependant on when the application of nanomaterials take place i.e. pre- or post- germination as well as different nanomaterials having different effects on plant species and organisms. So these factors i.e. size of nanoparticle, type of nanoparticle, when dispensed and plant species, all need to be considered when choosing a biological 'wire'.

\subsection{Implementation of logical circuits using plant-based memristors} 
\label{plantbasedmemristors}

Memristor (memory resistor) \index{memristor} is a device whose resistance changes depending on the polarity and magnitude of a voltage applied to the device's terminals and the duration of this voltage application. The memristor is a non-volatile memory because the specific resistance is retained until the application of another voltage~\cite{chua1971memristor, chua1974memristive, strukov2008missing}. A memristor implements a material implication of Boolean logic and thus any logical circuit can be constructed from memristors \cite{Borghetti}. 

In 2013, we discovered memristive behaviour of a living substrate, namely slime mould \emph{Physarum polycephalum}~\cite{gale2015slime} while Volkov and colleagues reported that some plants, like Venus flytrap, \emph{Mimosa pudica} and \emph{Aloe vera}, exhibit characteristics analogous to memristors pitch curves in the electrical current versus voltage profiles~\cite{volkov2014memristors}.  A strong hypothesis is that more likely all living and unmodified plants are `memristors' as well as living substrates, including slime mould~\cite{gale2015slime}, skin~\cite{martinsen2010memristance} and blood~\cite{kosta2011human}.

In our case, the basic device is an organic memristive system-element, composed of conducting polymer PANI, as described earlier, with a solid electrolyte heterojunction. Device conductivity is a function of ionic charge which is transferred through the heterojunction. Its application for the realization of adaptive circuits and systems, imitating synaptic learning, has been already demonstrated
\cite{berzina2007spectroscopic, erokhin2007non, erokhin2008electrochemically, erokhin2011material, erokhin2012organic}. By employing PANI coated plant roots as memristive  devices, novel memristors-based electronics will be investigated and designed aimed at exploiting the potential advantages of this device in advanced information processing circuits. Original circuit design methodologies could be addressed to exploit the memristor non-linear behaviour and its memory properties in novel computational networks and especially when aiming at the design of beyond von Neumann computing systems. 

In particular, memristor-based logic circuits open new pathways for the exploration of advanced computing architectures as promising alternatives to conventional integrated circuit technologies which are facing serious challenges related to continuous scaling \cite{ITRS}, \cite{Linn}, \cite{Pershin}. However, up to now no standard logic circuit design methodology exists \cite{Vourkas}. So, it is not immediately clear what kind of computing architectures would in practice benefit the most from the computing capabilities of memristors \cite{Borghetti,Gao,Shahar2012,VourkasS14,Shahar2014,Lehtonen2014,Papandroulidakis,Vourkas,Vourkasbook}.

Even if we wish to apply some hybrid designs like 1 transistor -- 1 memristor (1T1M) to further explore the computing paradigms of such structures, while still lying in the living substrate level, 
the plants can be modified using a similar approach developed in \cite{tarabella2015hybrid}. Tarabella \emph{et al.} implemented a transistor, a three-terminal active device that power amplifies an input signal, with slime mould. An organic electrochemical transistor is a semiconducting polymer channel in contact with an electrolyte. Its functioning is based on the reversible doping of the polymer channel.  A hybrid Physarum bio-organic electrochemical transistor was made by interfacing an organic semiconductor, poly-3,4-ethylenedioxythiophene doped with poly-styrene sulfonate, with the Physarum~\cite{tarabella2015hybrid}. The slime mould played a role of electrolyte. Electrical measurements in three-terminal mode uncover characteristics similar to transistor operations. The device operates in a depletion mode similarly to standard  electrolyte-gated transistors. The Physarum transistor works well with platinum, golden and silver electrodes. If the drain electrode is removed and the device becomes two-terminal, it exhibits cyclic voltage-current characteristics similar to memristors~\cite{tarabella2015hybrid}. We are aiming to apply similar confrontation in the plant devices and discover by experiments the limits of the proposed approach.

As a result, basic logic and analog functionalities will be scouted for the specific organic memristor plant devices and should be compared with standard implementations on silicon. Our focus would be on the inherent low voltage and low power consumption of the proposed devices and it would be a key issue of the work. This approach will hopefully allow us to construct all the necessary universal logical circuits either by using one of the possible already existing, beyond von Neumann, computing architectures like, for example, material implication or by introducing a novel suitable, beyond-Von Neumann, computing system architecture.

\section{Analog computation on electrical properties of plant roots}
\label{analogcomputation}

In tasks of collision-based computing and implementation of conservative logical gates, apexes of roots represented discrete quanta of signals while acting in continuous time. When implementing analog computing devices with roots we adopt continuous representation of space and time and continuous values. 

\index{analog computation}
The following components will be analysed: resistors, capacitors, operational amplifiers, multipliers, potentiometers and fixed-function generators. We will evaluate a potential towards implementation of the core mathematical operations that will be implemented in experimental laboratory conditions: summation, integration with respect to time, inversion, multiplication, exponentiation, logarithm, division. The mathematical and engineering problems to be solved will be represented in plant root networks of resistive elements or reaction elements, involving capacitance and inductance as well as resistance, to model spatial distribution of voltage, current, electrical potential in space, temperature, pressure \cite{peterson1967basic, johnson1963analog, weyrick1969fundamentals}. Implementations considered will included active elements analog computers, where no amplification required and passive elements computers, where amplification of signal is necessary. 

Typically data in analogue circuits can be represented by resistors. 
By selectively modifying properties some parts of plant's root system, we make the plant to implement basic analog computing~\cite{soroka1954analog, james1966analog}. A feasibility of constructing plant circuits with heterogeneous functionality of parts is supported by results showing that cerium dioxide nanoparticles in-taken by maize plants are not transferred to the newly grown areas of plants \cite{birbaum2010no}.
Let us consider an example of an analog circuit which can be produced from plants with modified electrical properties. This example is borrowed from our paper \cite{gizzie2016hybridising}.

\begin{figure}[!tbp]
\centering
\includegraphics[width=0.45\textwidth]{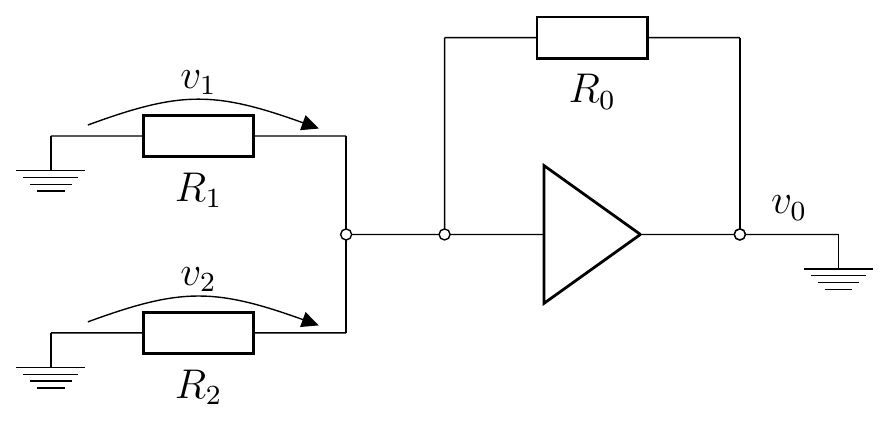}
    \caption{Summing amplifier simulating equation $z=ax+by$, $a=\frac{R_0}{R_1}$, 
    $b=\frac{R_0}{R_2}$, $v_1=-x$ and $v_2=-y$, $v_0 = z$; see discussion in \cite{gizzie2016hybridising}. }
  \label{summmingamplifier}
\end{figure}

If we selectively adjust resistance of two branches of the same root and interface upper part of the root with an amplifier the plant will simulate certain equations. The classical (see e.g. \cite{soroka1954analog, james1966analog}) scheme is shown in Fig.~\ref{summmingamplifier}. Root branches with modified electrical properties represent input resistors $R_1$ and $R_2$, upper modified root represents feedback resistance $R_0$. The output voltage $v_0$ is equal to the negative of the algebraic sums of the modified input voltages: 
$v_0 = - (\frac{R_0}{R_1}v_1 + \frac{R_0}{R_2}v_2)$. Depending on the ratios of the resistances and voltage polarity in each branch the scheme implements addition, subtraction and multiplication. The equation $v_0 = - (\frac{R_0}{R_1}v_1 + \frac{R_0}{R_2}v_2)$ can be seen as analogous to $z=ax+by$, where the output voltage $v_0$ has a polarity corresponding to the sign of $z$, the sign inversion of the amplifier is compensated by negative relationships between $v_1$ and $x$ and $v_2$ and $y$: $v_1=-x$ and $v_2=-y$. Thus, we have $a=\frac{R_0}{R_1}$ and $b=\frac{R_0}{R_2}$. 

Then we managed to selectively modify root branches of the same root with graphene $R_1=R_{\text{graphene}}\approx 2$M$\Omega$  
and calcium phosphate $R_2=R_{\text{CaPh}}\approx 4$M$\Omega$~\cite{gizzie2016hybridising}. Part of the root above branching site remains unmodified: $R_0=R_{\text{control}}\approx 3$M$\Omega$. Thus, the modified root simulates the equation $z=1.3 x + 2y$. The input voltages are multiplied by $1.3$ and $2.0$ and then added. If $a$ and $b$ are less than 1 then input voltage will be divided and then added: this can be achieved by, e.g. leaving one of the branches unmodified and loading part of the root above branching site with graphene. By making $v_1$ and $v_2$ with opposite polarity we simulate subtraction. To make a summing integration we need to substantially increase capacitance of some segments of a root~\cite{gizzie2016hybridising}.

The following tasks can be potentially implemented in plant-based analog computers:
\begin{itemize}
\item Hamilton circuit and Travelling Salesman Problem~\cite{ore1960note,bellman1962dynamic}. Given cities, physically represented by current differences in the plant root network, find a shortest path through all cities, which visits each city once and ends in its source. Associated potential, or Lyapunov, function achieves its global minimum value at an equilibrium point of the network corresponding to the problem's solution.
\item  Satisfiability Problem~\cite{kalmar1947reduction, dowling1984linear}. Values of variables of a Boolean logic formula are represented by voltage values of the root tree-like network and potential is indicated if the values can be assigned in a manner to make the Boolean formula true.
\item Quadratic Diophantine Equation and partition problem~\cite{adleman1994open}. Positive input numbers $a$, $b$, and $c$ are represented by sources of electrical current to answer the question --- are two positive integers $x$ and $y$ such that $(a \cdot x^2)+(b \cdot y)=c$.
\item Majority-classification (earliest cellular automaton version is published in \cite{gacs1978one}). Given a configuration of input states $a$ and $b$, represented by electrical characteristics of distant parts of plant root network, generate output value $a$ if majority of inputs have value $a$ and generated value $b$ if the value $b$ dominates in the inputs.
\item Analog sorting of numbers~\cite{akl2014parallel}. Usually sorting of numbers assume discreteness of data. We can represent values of numbers to be sorted by electrical currents and apply principles of rational flow in a non-periodic Toda lattice \cite{brockett1997rational} to undertake the smooth sorting.
\item Implementation of Kirchhoff-{\L}ukasiewicz machine~\cite{mills1995kirchhoff,mills2008nature}.
A Kirchhoff-{\L}ukasi\-ewicz machine was proposed by late Jonathan Mills
\cite{mills1995kirchhoff,mills2008nature} to combine power and intuitive appeal of analog computers with conventional digital circuits. The machine is based on a sheet of conductive foam with array of probes interfaced with hardware implementation of {\L}ukasiewcz logic arrays. The {\L}ukasiewcz arrays are regular lattices of continuous state machines connected locally to each other. Arithmetic and logical functions are defined using implication and negated implication. Array inputs are differences between two electrical currents.  Algebraic expressions can be converted to {\L}ukasiewicz implications by tree-pattern matching and minimisation. The simplified expressions of implications can be further converted to layouts of living and mineralised/coated plant roots. The tasks to be implemented on plant-root based Kirchoff-{\L}ukasiewicz machine are fuzzy controllers, tautology checkers, simulation of constraint propagation network with implications.
\end{itemize}

\section{Evolution in plants: searching for logical gates}
\index{logical gate}

Given the success of implementing logical gates with the Mecobo evolvable hardware platform and Physarum~\cite{harding2016discovering}, it was envisioned that plants may also be a suitable medium for `evolution in materio'\index{evolution in materio} type experiments. The Mecobo was designed to provide a general purpose interface to allow for evolution in materio experiments and for probing the electrical properties of a substrate without understanding the underlying electrical properties of the substrate, and without having to develop new interface techniques for each material under investigation \cite{lykkebo2014mecobo}. 

The same methodology was used to find gates as had previously been used for Physarum~\cite{harding2016discovering}, and carbon nano-tubes and various polymers \cite{miller2014evolution,lykkebo2015investigation,massey2015computing}.

\begin{figure}[!tbp]
\centering
\subfigure[]{\includegraphics[width=0.48\textwidth]{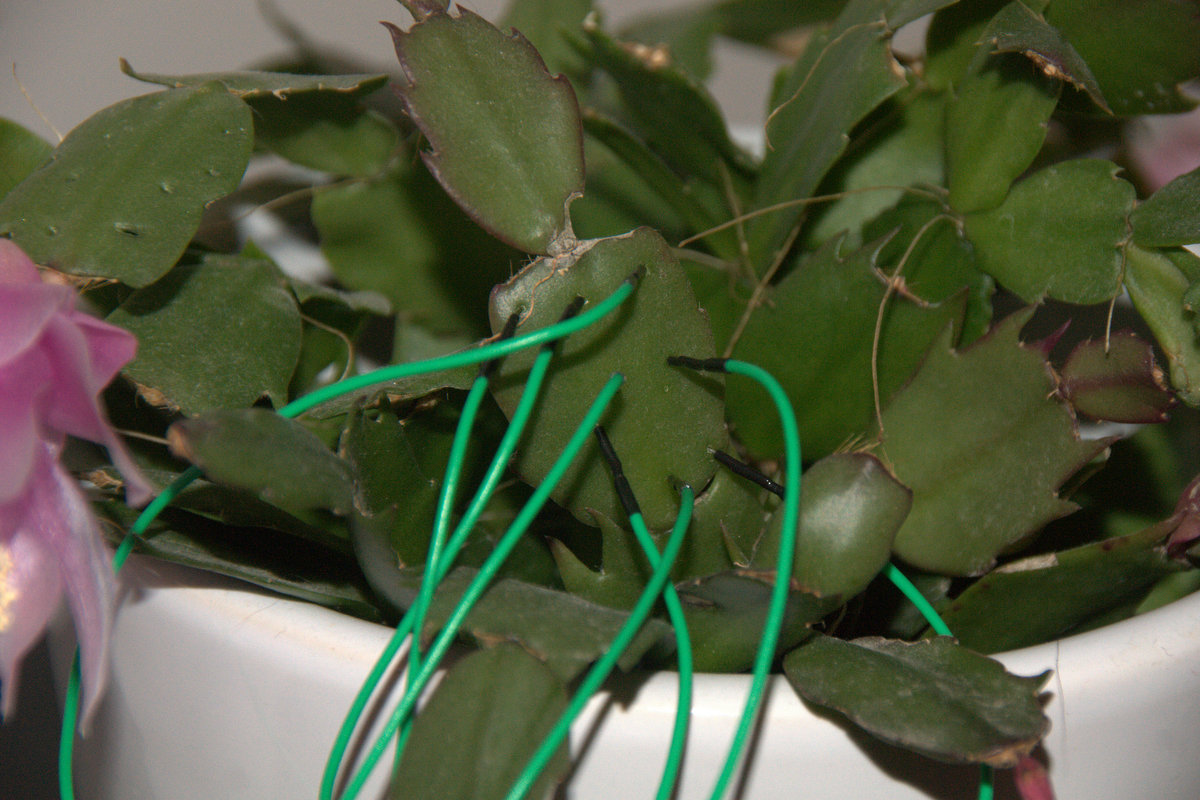}}
\subfigure[]{\includegraphics[width=0.48\textwidth]{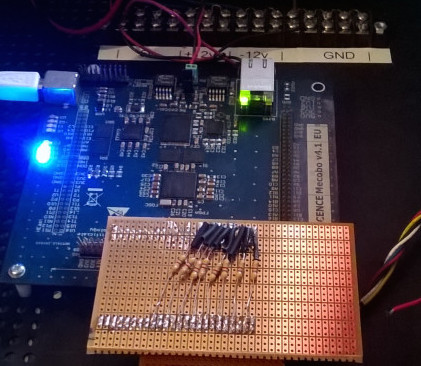}}
\caption{Photos of experimental setup.
(a)~Electrodes in Schlumbergera cactus.
(b)~Mecobo board.}
\label{cactus}
\end{figure}

Eight electrodes were connected to the digital outputs of the Mecobo, via a current limiting 4.7\,k$\Omega$ resistor, as shown in Fig.~\ref{cactus}. The electrodes were then inserted through the stem of a plant. For these experiments, a common house hold plant, Schlumbergera, was used. Schlumbergera is a type of Brazilian cactus commonly known as a `Christmas Cactus'. It has large, flat, stem segments that provide a good location to securely insert electrodes. All of the electrodes were inserted into a single segment. The positions chosen were essentially random, but with care taken to ensure that two electrodes were not very close together or touching as this would likely result in electrodes being shorted together, and reducing the chances of finding interesting behaviour. An example arrangement can be seen in Fig.~\ref{cactus}a.

Similar to before, an exhaustive search was conducted by applying all possible binary combinations of various frequency pairs to 7 pins (which is a practical amount for time purposes).  One pin was used as an output from the material, with the other 7 pins acting as inputs to the plant.

For each binary combination, each pair of frequencies was tried with one frequency representing a `low' and the other representing a `high' input. The frequency pairs were combinations of DC square waves of either $(250$\,Hz, $500$\,Hz, $1$\,kHz, or $2.5$\,kHz$)$. The amplitude of the square waves is $3.3$V.

\begin{table}[!tbp]
\begin{center}
\caption{Number of gates mined from the frequency responses of the Schlumbergera
}
\label{tab:Schlumbergera:AllGatesAllRuns}
\begin{tabular}{ccccccccc}
\toprule
{\bf Cfg.}& \multicolumn{4}{c}{Inputs $xy$} & Number of gates &{\bf Gate} \\
   & FF & FT & TF & TT & & &  \\
\midrule
1	& F & F&F	& F & 95718	& Constant False\\					
2	& T & F&F	& F & 366	& 	$x$	NOR	$y$	\\			
3	& F & T&F	& F & 304	& 	NOT	$x$	AND	$y$\\			
4	& T & T&F	& F & 430	& 	NOT	$x$\\					
5	& F & F&T	& F & 304	& 	$x$	AND	NOT	$y$\\			
6	& T & F&T	& F & 430	& 	NOT	$y$\\					
7	& F & T&T	& F & 74	& 	$x$	XOR	$y$	\\			
8	& T & T&T	& F & 314	& 	$x$	NAND	$y$	\\			
9	& F & F&F	& T & 510	& 	$x$	AND	$y$		\\		
10	& T & F&F	& T & 104	& 	$x$	XNOR	$y$	\\			
11	& F & T&F	& T & 863	& 	$y$\\						
12	& T & T&F	& T & 307	& 	NOT	$x$	AND	NOT	$y$	OR	$y$\\
13	& F & F&T	& T & 863	& 	$x$\\						
14	& T & F&T	& T & 307	& 	$x$	OR	NOT	$y$\\			
15	& F & T&T	& T & 512	& 	$x$	OR	$y$	\\			
16	& T & T&T	& T & 94564	& Constant True\\				
\bottomrule
\end{tabular}
\addtolength{\tabcolsep}{3pt}
\end{center}
\end{table}

\begin{table}[!tbp]
\begin{center}
\caption{Number of gates mined from the frequency responses of the Schlumbergera
}
\label{tab:Schlumbergera:AllGatesByRun}
\begin{tabular}{cccccccc}
\toprule
{\bf Cfg.} & \multicolumn{6}{c}{Number of gates} &{\bf Gate} \\
   &  Run 1 &  Run 2 & Run 3 & Run 4 & Run 5 & Average &  \\
\midrule
1& 19130& 18870& 	19160 &	19198 &	19360	 &19144  & Constant False\\					
2& 56	 &50	 &104& 90& 66& 73     & 	$x$	NOR	$y$	\\			
3& 56	 &42	 &64& 67& 75& 61    & 	NOT	$x$	AND	$y$\\			
4& 99	 &72	 &79& 89& 91& 86     & 	NOT	$x$\\					
5& 56	 &42	 &64& 67& 75& 61     & 	$x$	AND	NOT	$y$\\			
6& 99	 &72	 &79& 89& 91& 86 & 	NOT	$y$\\					
7& 4	 & 12	 &20	 &22	 &16& 15     & 	$x$	XOR	$y$	\\			
8& 68	 &68	 &68	& 52	& 58	& 63      & 	$x$	NAND	$y$	\\			
9& 88	 &70	 &114	& 118	& 120	& 102  & 	$x$	AND	$y$		\\		
10& 8	& 8	 &38	 &32	& 18	& 21    &   	$x$	XNOR	$y$	\\			
11& 89	 &71	 &243	& 228	& 232	& 173 & 	$y$\\						
12& 57	 &52	 &63	& 60	& 75	& 61 & 	NOT	$x$	AND	NOT	$y$	OR	$y$\\
13& 89	 &71	 &243	& 228	& 232	& 173  & 	$x$\\						
14& 57	 &52	 &63	& 60	& 75	& 61  & 	$x$	OR	NOT	$y$\\			
15& 68	 &44	 &138	& 136	& 126	& 102  & 	$x$	OR	$y$	\\			
16& 19134& 18844& 	18840 &	18842 &18904	 &18913  & Constant True\\				

\bottomrule
\end{tabular}
\addtolength{\tabcolsep}{3pt}
\end{center}
\end{table}

The Mecobo measured the digital response from the plant for 32ms. The digital threshold is $0.75$V for high, with voltages below this being classed as low. The sampling frequency was twice the highest input frequency applied.

Five different runs were completed. With different stem segments and electrode arrangements used each time.

Table~\ref{tab:Schlumbergera:AllGatesAllRuns}, shows a summary for all of the runs. We see that all possible 2 input Boolean gates were implemented. As with previous work, we see that gates such as XOR and XNOR are found relatively infrequently. Looking at each run individually, Table~\ref{tab:Schlumbergera:AllGatesByRun} shows the same pattern. It is interesting to note that on each run all gates were found, and that in very similar proportions. 

Table~\ref{tab:Schlumbergera:FrequencyForXor} shows how many XOR gates were found for each combination of frequencies used for representing the Boolean input states. We see that there is a bias towards the higher frequencies tested. We also see that not all combinations produce gates, and that the frequency pairs are not used asymmetrically. For example, true and false can be represented by either 2500Hz or 1000Hz, but representing false with 2500Hz produces more viable gates. It appears that representing false by the higher frequency in the pair produces more solutions. More in depth analysis, and modelling of the results will be required to fully understand this behaviour, but it hints that expanding the search to use higher frequencies than 2500Hz would result in more Boolean circuits being discovered.

\begin{table}[!tbp]
\begin{center}
\caption{Number of XOR gates mined from the frequency responses of the Schlumbergera for each frequency pair. Frequency A is used to represent False, Frequency B for True. 
}
\label{tab:Schlumbergera:FrequencyForXor}
\begin{tabular}{ccc}
\toprule
{\bf Frequency A} & {\bf Frequency B}  &{\bf Count}  \\
\midrule
2500  & 1000 & 	28\\
1000  & 250	 & 14\\
1000 &  2500 & 	10\\
250  & 2500	 & 8\\
500 &  250 & 	8\\
2500  & 500	 & 4\\
2500  & 250 & 	2\\
1000 &  500 & 0 \\
500 &  2500 & 0 \\
500 &  1000 & 0 \\
250  & 1000 & 0 \\
250 &  500 & 0 \\
\bottomrule
\end{tabular}
\addtolength{\tabcolsep}{3pt}
\end{center}
\end{table}

Whilst the experiments with Schlumbergera were successful, experiments with other plants will be necessary to prove feasibility of the approach. Chances are high plants of different species will be showing differently shaped distributions of frequencies of logical gates discovered. Thus, we might construct  a unique mapping between taxonomy of plants and geometries of logical gates distributions. Also, methodology wise, future experiments will investigate if it is possible to find solutions directly using evolution to find the pin configuration, rather than a time consuming exhaustive approach.

\section{Brain made of plants}
\label{brainplants}

\begin{figure}[!tbp]
\centering
\subfigure[]{\includegraphics[width=0.55\textwidth]{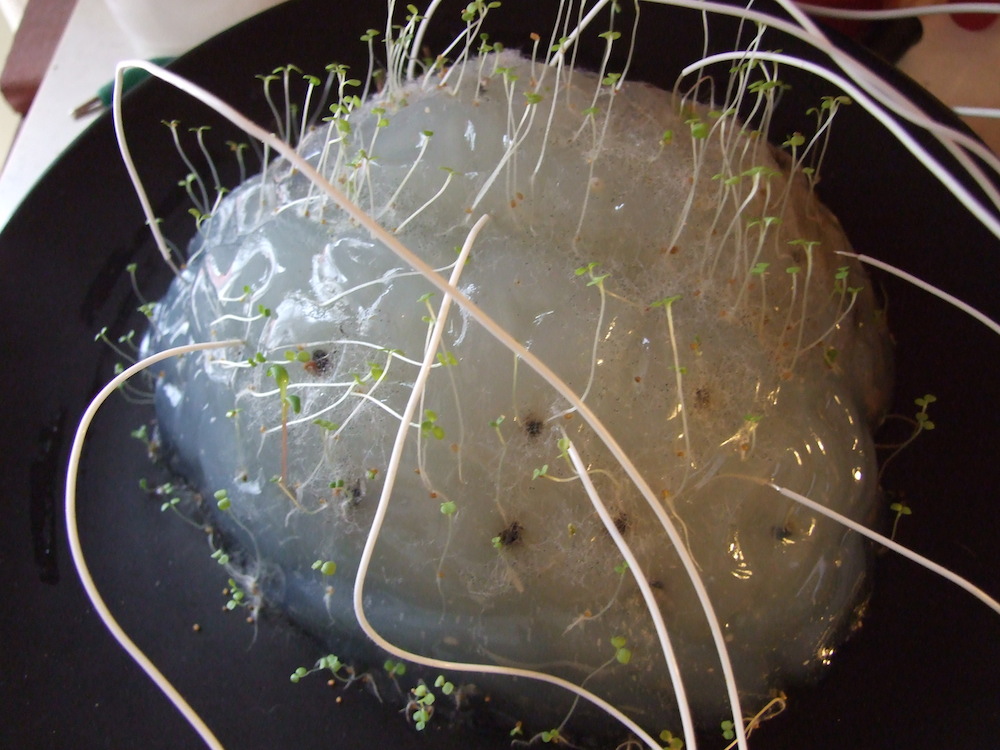}}
\subfigure[]{\includegraphics[width=0.44\textwidth]{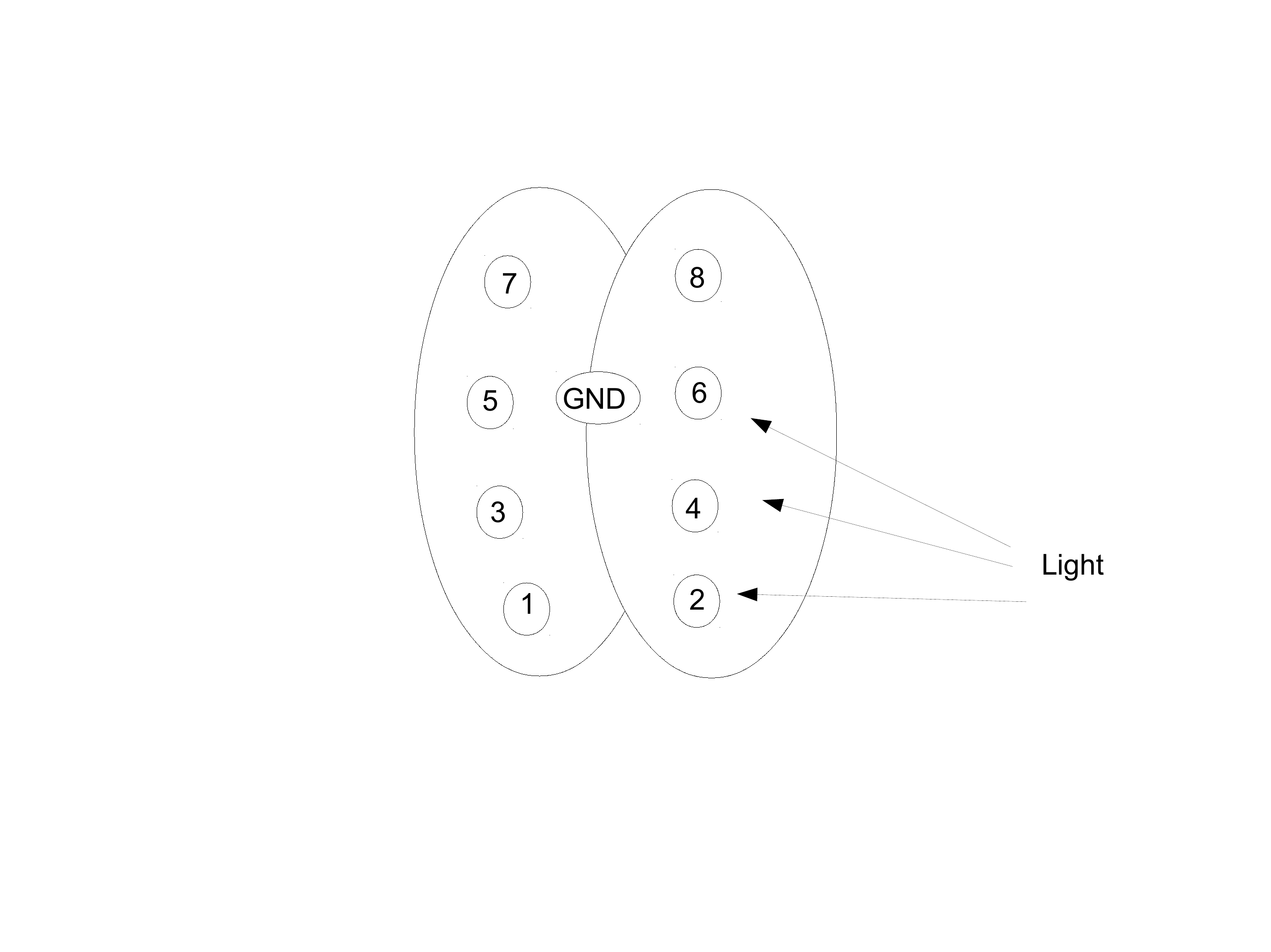}}
\subfigure[]{\includegraphics[width=0.9\textwidth]{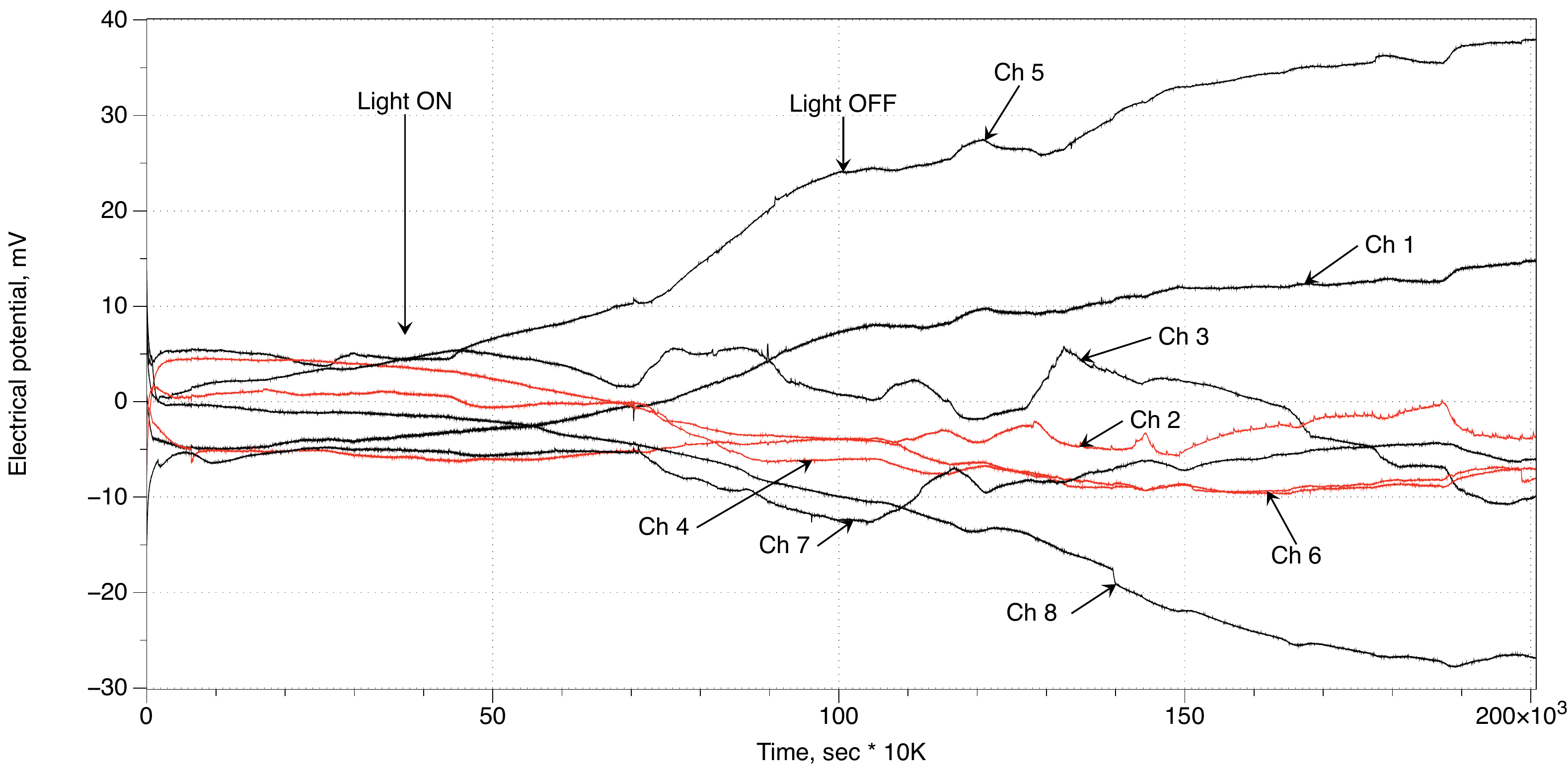}}
\caption{Experimental setup towards a `brain' made of lettuce seedlings. 
(a)~A template of human cortex made from phytogel, lettuce seedlings and electrodes are visible. 
(b)~Scheme of positions of electrodes and site of stimulation with light.
(c)~Electrical potential recorded over two days with sampling rate once per second.
}
\label{experimentbrain}
\end{figure}

The survival of an organism depends on its ability to respond to the environment through its senses when neuronal systems translate sensory information into electrical impulses via neural code \cite{deweese2006neurobiology}. This enables multisensory integration that in turn leads to adaptive motor responses and active behavior. In plants, numerous physical environmental factors, especially light and gravity, are continuously monitored
\cite{baluvska2005plant, baluvska2006communication, baluvska2006neurobiological, gagliano2012towards, gagliano2014experience, mancuso1999hydraulic, stone2005phototropins, brenner2006plant}. Specialized plant cells have been optimized by evolution to translate sensory information obtained from the physical environment into motor responses known as tropisms, with root gravitropism representing one of the most intensively studied plant organ tropisms. Electrical signals are induced by all known physical factors in plants, suggesting that electricity mediates physical-biological communication in plants 
too~\cite{baluvska2006communication, volkov2006plants, volkov2006electrophysiology, fromm2007electrical, felle2007systemic}. Root gravitropism is a particularly instructive example in this respect. Sensory perception of gravity is accomplished at the very tip of the root apex, at the root cap~\cite{barlow1993response}, which includes a vestibular-like gravisensing organ composed of statocytes~\cite{baluvska2009deep}. On the other hand, the motor responses, which begin almost immediately after root cap sensory events, are accomplished in relatively remote growth zones of the root apex~\cite{baluvska2007plant}. Therefore, root gravitropism represents a nice example of a neuronal sensory-motor circuit in plants. 

Quite possible future green computers will be using plants as elements of memory and neural network, followed by plant neuromorphic processors, will be developed. A rather artistic example of an attempt to make a neural-like network from plants is shown in Fig.~\ref{experimentbrain}. Seeds of lettuce were planted into the phytogel-made real-life sized model of a human brain cortex. When seedlings developed (Fig.~\ref{experimentbrain}a) recording electrodes were placed in the model brain (Fig.~\ref{experimentbrain}b); reference electrode was located medially between hemispheres. Electrical potential on each recording electrode was determined by electrical potential of c. 20 lettuce seedlings in the vicinity of the electrode.

To check if there is an interaction between seedlings in a response to stimulation we illuminated a group of seedlings. Electrical response of the lettuce population is shown in Fig.~\ref{experimentbrain}c. Indeed one experiment must not lead to any conclusion, however, we can still speculate that light stimulus might lead to slight decrease in electrical potential of the illuminated population. The light stimulation might also lead to wider dispersion of electrical potential values in the population.

\section{Discussion}
\label{discussion}

We proposed designs of living and hybrid computing systems made of plants. 
The designs are original both in theory --- collision-based computing, morphological computing, memristive circuits --- and implementation --- functional devices to be made of living or loaded with nanoparticles or coated with conductive polymers plant roots. When the proposed designs will be implemented they will contribute towards breakthroughs in computer science (algorithm/architectures of plant computing), computational geometry (plant based processors), graph-theoretic studies (proximity graphs by plant roots), biology (properties of plant at the interface with electronics), material science (functional elements), electronics (high density self-growing circuits), self-assembly and self-regenerative systems.
The designs can be materialised in two type of phyto-chips: morphological chip and analog chips. The morphological chips are disposable computing devices based on computation with raw, unmodified, living roots; these chips represent result of computation by geometry of grown parts of plants. Computational geometry processors will be solving plane tessellations and generalised Voronoi diagrams, approximation of planar shapes (concave and convex hulls); these are classical problems of computational geometry. Almost all classical tasks of image processing, including dilation, erosion, opening and closing, expansion and shrinking, edge detection and completion can be solved using swarms of plant roots.  Graph-optimisation processors will be solving tasks of parameterised spanning trees, Gabriel graph, relative neighbourhood graph, Delaunay triangulation, Steiner trees.
Analog chips are based on biomorphic mineralisation of plant root networks and coating roots with metals and alloys; some architecture could be composed of living and artificially functionalised root networks. The analog phyto-chips are general purpose analog computers, capable for implementing multi-valued logic and arithmetic.
What species of plants will be used in future phyto-chips? Best candidates are \emph{Arabidopsis thaliana}, \emph{Zea mays}, \emph{Ocimum basilicum}, \emph{Mentha genus}, \emph{Raphanus sativus}, \emph{Spinacia oleracea}, \emph{Macleaya microcarpa}, \emph{Helianthus annuus}. They could be chosen due to their growth characteristics, robustness,  availability of mutant lines and research data on physiology and protocols of laboratory experiments.


 \bibliographystyle{plain}
\bibliography{plantbib}

\begin{thebibliography}{100}

\bibitem{adamatzky2014JBIS}
A.~{Adamatzky}, R.~{Armstrong}, B.~{De Lacy Costello}, Y.~{Deng}, J.~{Jones},
  R.~{Mayne}, T.~{Schubert}, G.~C. {Sirakoulis}, and X.~{Zhang}.
\newblock {Slime Mould Analogue Models of Space Exploration and Planet
  Colonisation}.
\newblock {\em Journal of the British Interplanetary Society}, 67:290--304,
  2014.

\bibitem{adamatzky2002collision}
Andrew Adamatzky.
\newblock {\em Collision-based computing}.
\newblock Springer, 2002.

\bibitem{adamatzky2009hot}
Andrew Adamatzky.
\newblock Hot ice computer.
\newblock {\em Physics Letters A}, 374(2):264--271, 2009.

\bibitem{adamatzky2010physarum}
Andrew Adamatzky.
\newblock {\em Physarum {M}achines: {C}omputers from slime mould}, volume~74.
\newblock World Scientific, 2010.

\bibitem{adamatzky2012slimehull}
Andrew Adamatzky.
\newblock Slime mould computes planar shapes.
\newblock {\em International Journal of Bio-Inspired Computation},
  4(3):149--154, 2012.

\bibitem{adamatzky2014towards}
Andrew Adamatzky.
\newblock Towards plant wires.
\newblock {\em Biosystems}, 122:1--6, 2014.

\bibitem{adamatzky2016advances}
Andrew Adamatzky, editor.
\newblock {\em Advances in Physarum machines: Sensing and computing with slime
  mould}.
\newblock Springer, 2016.

\bibitem{adamatzky2005reaction}
Andrew Adamatzky, Benjamin De~Lacy Costello, and Tetsuya Asai.
\newblock {\em Reaction-{D}iffusion {C}omputers}.
\newblock Elsevier, 2005.

\bibitem{adamatzky2011computing}
Andrew Adamatzky, Julian Holley, Larry Bull, and Ben De~Lacy Costello.
\newblock On computing in fine-grained compartmentalised belousov--zhabotinsky
  medium.
\newblock {\em Chaos, Solitons \& Fractals}, 44(10):779--790, 2011.

\bibitem{adamatzky2011pluscomputing}
Andrew Adamatzky, Stephen Kitson, Ben De~Lacy Costello, Mario~Ariosto Matranga,
  and Daniel Younger.
\newblock Computing with liquid crystal fingers: Models of geometric and
  logical computation.
\newblock {\em Physical Review E}, 84(6):061702, 2011.

\bibitem{adamatzky2016plant}
Andrew Adamatzky, Georgios Sirakoulis, Genaro~J Martinez, Frantisek Baluska,
  and Stefano Mancuso.
\newblock On plant roots logical gates.
\newblock {\em arXiv preprint arXiv:1610.04602}, 2016.

\bibitem{adleman1994open}
Leonard~M Adleman and Kevin~S McCurley.
\newblock Open problems in number theoretic complexity, ii.
\newblock In {\em International Algorithmic Number Theory Symposium}, pages
  291--322. Springer, 1994.

\bibitem{akl2014parallel}
Selim~G Akl.
\newblock {\em Parallel sorting algorithms}, volume~12.
\newblock Academic press, 2014.

\bibitem{bais2006role}
Harsh~P Bais, Tiffany~L Weir, Laura~G Perry, Simon Gilroy, and Jorge~M Vivanco.
\newblock The role of root exudates in rhizosphere interactions with plants and
  other organisms.
\newblock {\em Annu. Rev. Plant Biol.}, 57:233--266, 2006.

\bibitem{bais2004plants}
Harsh~Pal Bais, Sang-Wook Park, Tiffany~L Weir, Ragan~M Callaway, and Jorge~M
  Vivanco.
\newblock How plants communicate using the underground information
  superhighway.
\newblock {\em Trends in plant science}, 9(1):26--32, 2004.

\bibitem{baluvska2007plant}
Franti{\v{s}}ek Balu{\v{s}}ka and Stefano Mancuso.
\newblock Plant neurobiology as a paradigm shift not only in the plant
  sciences.
\newblock {\em Plant signaling \& behavior}, 2(4):205--207, 2007.

\bibitem{baluvska2009deep}
Franti{\v{s}}ek Balu{\v{s}}ka and Stefano Mancuso.
\newblock Deep evolutionary origins of neurobiology: turning the essence
  of'neural'upside-down.
\newblock {\em Communicative \& integrative biology}, 2(1):60--65, 2009.

\bibitem{baluvska2009plant}
Franti{\v{s}}ek Balu{\v{s}}ka and Stefano Mancuso.
\newblock Plant neurobiology: from sensory biology, via plant communication, to
  social plant behavior.
\newblock {\em Cognitive processing}, 10(1):3--7, 2009.

\bibitem{baluvska2016vision}
Frantisek Balu{\v{s}}ka and Stefano Mancuso.
\newblock Vision in plants via plant-specific ocelli?
\newblock {\em Trends in Plant Science}, 21(9):727--730, 2016.

\bibitem{plantsignallingbook}
Frantisek Baluska, Stefano Mancuso, and Dieter Voljmann, editors.
\newblock {\em Communication in Plants: Neuronal Aspects of Plant Life}.
\newblock Springer, 2007.

\bibitem{baluvska2006communication}
Franti{\v{s}}ek Balu{\v{s}}ka, Stefano Mancuso, and Dieter Volkmann.
\newblock Communication in plants.
\newblock {\em Neuronal aspect of plant life., Spriger Berlin Heidelberg, New
  York}, 2006.

\bibitem{baluvska2004root}
Franti{\v{s}}ek Balu{\v{s}}ka, Stefano Mancuso, Dieter Volkmann, and Peter
  Barlow.
\newblock Root apices as plant command centres: the unique
  ‘brain-like’status of the root apex transition zone.
\newblock {\em Biologia (Bratisl.)}, 59(Suppl. 13):1--13, 2004.

\bibitem{baluvska2010root}
Franti{\v{s}}ek Balu{\v{s}}ka, Stefano Mancuso, Dieter Volkmann, and Peter~W
  Barlow.
\newblock Root apex transition zone: a signalling--response nexus in the root.
\newblock {\em Trends in plant science}, 15(7):402--408, 2010.

\bibitem{baluvska2006neurobiological}
Franti{\v{s}}ek Balu{\v{s}}ka, Dieter Volkmann, Andrej Hlavacka, Stefano
  Mancuso, and Peter~W Barlow.
\newblock Neurobiological view of plants and their body plan.
\newblock In {\em Communication in plants}, pages 19--35. Springer, 2006.

\bibitem{baluvska2005plant}
Franti{\v{s}}ek Balu{\v{s}}ka, Dieter Volkmann, and Diedrik Menzel.
\newblock Plant synapses: actin-based domains for cell-to-cell communication.
\newblock {\em Trends in plant science}, 10(3):106--111, 2005.

\bibitem{barlow1993response}
Peter~W Barlow.
\newblock The response of roots and root systems to their environment—an
  interpretation derived from an analysis of the hierarchical organization of
  plant life.
\newblock {\em Environmental and experimental botany}, 33(1):1--10, 1993.

\bibitem{battistoni2016spectrophotometric}
S~Battistoni, A~Dimonte, and V~Erokhin.
\newblock Spectrophotometric characterization of organic memristive devices.
\newblock {\em Organic Electronics}, 38:79--83, 2016.

\bibitem{battistoni2017organic}
Silvia Battistoni, Alice Dimonte, and Victor Erokhin.
\newblock Organic memristor based elements for bio-inspired computing.
\newblock In {\em Advances in Unconventional Computing}, pages 469--496.
  Springer, 2017.

\bibitem{bellman1962dynamic}
Richard Bellman.
\newblock Dynamic programming treatment of the travelling salesman problem.
\newblock {\em Journal of the ACM (JACM)}, 9(1):61--63, 1962.

\bibitem{berzina2007spectroscopic}
Tatiana Berzina, Victor Erokhin, and MP~Fontana.
\newblock Spectroscopic investigation of an electrochemically controlled
  conducting polymer-solid electrolyte junction.
\newblock {\em Journal of applied physics}, 101(2):024501, 2007.

\bibitem{birbaum2010no}
Karin Birbaum, Robert Brogioli, Maya Schellenberg, Enrico Martinoia, Wendelin~J
  Stark, Detlef G{\"u}nther, and Ludwig~K Limbach.
\newblock No evidence for cerium dioxide nanoparticle translocation in maize
  plants.
\newblock {\em Environmental science \& technology}, 44(22):8718--8723, 2010.

\bibitem{Borghetti}
J.~Borghetti, G.~S. Snidera, P.~J. Kuekes, J.~J. Yang, D.~R. Stewart, and R.~S.
  Williams.
\newblock {‘Memristive’ switches enable ‘stateful’ logic operations via
  material implication}.
\newblock {\em Nature}, 464(7290):873--–876, April 2010.

\bibitem{brenner2006plant}
Eric~D Brenner, Rainer Stahlberg, Stefano Mancuso, Jorge Vivanco,
  Franti{\v{s}}ek Balu{\v{s}}ka, and Elizabeth Van~Volkenburgh.
\newblock Plant neurobiology: an integrated view of plant signaling.
\newblock {\em Trends in plant science}, 11(8):413--419, 2006.

\bibitem{brockett1997rational}
Roger~W Brockett.
\newblock A rational flow for the toda lattice equations.
\newblock In {\em Operators, Systems and Linear Algebra}, pages 33--44.
  Springer, 1997.

\bibitem{burbach2012photophobic}
Christian Burbach, Katharina Markus, Yin Zhang, Markus Schlicht, and
  Franti{\v{s}}ek Balu{\v{s}}ka.
\newblock Photophobic behavior of maize roots.
\newblock {\em Plant signaling \& behavior}, 7(7):874--878, 2012.

\bibitem{chua1971memristor}
Leon Chua.
\newblock Memristor --- the missing circuit element.
\newblock {\em IEEE Transactions on circuit theory}, 18(5):507--519, 1971.

\bibitem{chua1974memristive}
Leon~O Chua and Chong-Wei Tseng.
\newblock A memristive circuit model for p-n junction diodes.
\newblock {\em International Journal of Circuit Theory and Applications},
  2(4):367--389, 1974.

\bibitem{cifarelli2015bio}
Angelica Cifarelli, Tatiana Berzina, and Victor Erokhin.
\newblock Bio-organic memristive device: polyaniline--physarum polycephalum
  interface.
\newblock {\em physica status solidi (c)}, 12(1-2):218--221, 2015.

\bibitem{ciszak2012swarming}
Marzena Ciszak, Diego Comparini, Barbara Mazzolai, Frantisek Baluska, F~Tito
  Arecchi, Tam{\'a}s Vicsek, and Stefano Mancuso.
\newblock Swarming behavior in plant roots.
\newblock {\em PLoS One}, 7(1):e29759, 2012.

\bibitem{costello2011towards}
Ben De~Lacy Costello, Andy Adamatzky, Ishrat Jahan, and Liang Zhang.
\newblock Towards constructing one-bit binary adder in excitable chemical
  medium.
\newblock {\em Chemical Physics}, 381(1):88--99, 2011.

\bibitem{costello2005experimental}
Benjamin De~Lacy Costello and Andrew Adamatzky.
\newblock Experimental implementation of collision-based gates in
  belousov--zhabotinsky medium.
\newblock {\em Chaos, Solitons \& Fractals}, 25(3):535--544, 2005.

\bibitem{demin2015hardware}
VA~Demin, VV~Erokhin, AV~Emelyanov, S~Battistoni, G~Baldi, S~Iannotta,
  PK~Kashkarov, and MV~Kovalchuk.
\newblock Hardware elementary perceptron based on polyaniline memristive
  devices.
\newblock {\em Organic Electronics}, 25:16--20, 2015.

\bibitem{deweese2006neurobiology}
Michael~R DeWeese and Anthony Zador.
\newblock Neurobiology: efficiency measures.
\newblock {\em Nature}, 439(7079):920--921, 2006.

\bibitem{dimonte2016physarum}
Alice Dimonte, Silvia Battistoni, and Victor Erokhin.
\newblock Physarum in hybrid electronic devices.
\newblock In {\em Advances in Physarum Machines}, pages 91--107. Springer,
  2016.

\bibitem{dowling1984linear}
William~F Dowling and Jean~H Gallier.
\newblock Linear-time algorithms for testing the satisfiability of
  propositional horn formulae.
\newblock {\em The Journal of Logic Programming}, 1(3):267--284, 1984.

\bibitem{emelyanov2016first}
AV~Emelyanov, DA~Lapkin, VA~Demin, VV~Erokhin, S~Battistoni, G~Baldi,
  A~Dimonte, AN~Korovin, S~Iannotta, PK~Kashkarov, et~al.
\newblock First steps towards the realization of a double layer perceptron
  based on organic memristive devices.
\newblock {\em AIP Advances}, 6(11):111301, 2016.

\bibitem{erokhin2007non}
Victor Erokhin, Tatiana Berzina, Paolo Camorani, and Marco~P Fontana.
\newblock Non-equilibrium electrical behaviour of polymeric electrochemical
  junctions.
\newblock {\em Journal of Physics: Condensed Matter}, 19(20):205111, 2007.

\bibitem{erokhin2011material}
Victor Erokhin, Tatiana Berzina, Paolo Camorani, Anteo Smerieri, Dimitris
  Vavoulis, Jianfeng Feng, and Marco~P Fontana.
\newblock Material memristive device circuits with synaptic plasticity:
  learning and memory.
\newblock {\em BioNanoScience}, 1(1-2):24--30, 2011.

\bibitem{erokhin2008electrochemically}
Victor Erokhin and Marco~P Fontana.
\newblock Electrochemically controlled polymeric device: a memristor (and more)
  found two years ago.
\newblock {\em arXiv preprint arXiv:0807.0333}, 2008.

\bibitem{erokhin2012organic}
Victor Erokhin, Gerard~David Howard, and Andrew Adamatzky.
\newblock Organic memristor devices for logic elements with memory.
\newblock {\em International Journal of Bifurcation and Chaos}, 22(11):1250283,
  2012.

\bibitem{felle2007systemic}
Hubert~H Felle and Matthias~R Zimmermann.
\newblock Systemic signalling in barley through action potentials.
\newblock {\em Planta}, 226(1):203--214, 2007.

\bibitem{fredkin2002conservative}
Edward Fredkin and Tommaso Toffoli.
\newblock Conservative logic.
\newblock In Andrew Adamatzky, editor, {\em Collision-Based Computing}.
  Springer, 2002.

\bibitem{fromm2007electrical}
J{\"o}rg Fromm and Silke Lautner.
\newblock Electrical signals and their physiological significance in plants.
\newblock {\em Plant, Cell \& Environment}, 30(3):249--257, 2007.

\bibitem{gacs1978one}
P{\'e}ter G{\'a}cs, Georgy~L Kurdyumov, and Leonid~A Levin.
\newblock One-dimensional uniform arrays that wash out finite islands.
\newblock {\em Problemy Peredachi Informatsii}, 14(3):92--96, 1978.

\bibitem{gagliano2012towards}
Monica Gagliano, Stefano Mancuso, and Daniel Robert.
\newblock Towards understanding plant bioacoustics.
\newblock {\em Trends in plant science}, 17(6):323--325, 2012.

\bibitem{gagliano2014experience}
Monica Gagliano, Michael Renton, Martial Depczynski, and Stefano Mancuso.
\newblock Experience teaches plants to learn faster and forget slower in
  environments where it matters.
\newblock {\em Oecologia}, 175(1):63--72, 2014.

\bibitem{gagliano2012acoustic}
Monica Gagliano, Michael Renton, Nili Duvdevani, Matthew Timmins, and Stefano
  Mancuso.
\newblock Acoustic and magnetic communication in plants: is it possible?
\newblock {\em Plant signaling \& behavior}, 7(10):1346--1348, 2012.

\bibitem{gale2015slime}
Ella Gale, Andrew Adamatzky, and Ben de~Lacy~Costello.
\newblock c.
\newblock {\em BioNanoScience}, 5(1):1--8, 2015.

\bibitem{Gao}
L.~Gao, F.~Alibart, and D.~B. Strukov.
\newblock Programmable cmos/memristor threshold logic.
\newblock {\em IEEE Transactions on Nanotechnology}, 12(2):115--119, March
  2013.

\bibitem{geddes1967specific}
LA~Geddes and LE~Baker.
\newblock The specific resistance of biological material—a compendium of data
  for the biomedical engineer and physiologist.
\newblock {\em Medical and biological engineering}, 5(3):271--293, 1967.

\bibitem{gizzie2016hybridising}
Nina Gizzie, Richard Mayne, David Patton, Paul Kendrick, and Andrew Adamatzky.
\newblock On hybridising lettuce seedlings with nanoparticles and the resultant
  effects on the organisms’ electrical characteristics.
\newblock {\em Biosystems}, 147:28--34, 2016.

\bibitem{graham1991flavonoid}
Terrence~L Graham.
\newblock Flavonoid and isoflavonoid distribution in developing soybean
  seedling tissues and in seed and root exudates.
\newblock {\em Plant physiology}, 95(2):594--603, 1991.

\bibitem{gunji2011robust}
Yukio-Pegio Gunji, Yuta Nishiyama, Andrew Adamatzky, Theodore~E Simos, George
  Psihoyios, Ch~Tsitouras, and Zacharias Anastassi.
\newblock Robust soldier crab ball gate.
\newblock {\em Complex systems}, 20(2):93, 2011.

\bibitem{harding2016discovering}
Simon Harding, Jan Koutnik, Klaus Greff, Jurgen Schmidhuber, and Andy
  Adamatzky.
\newblock Discovering boolean gates in slime mould.
\newblock {\em arXiv preprint arXiv:1607.02168}, 2016.

\bibitem{james1966analog}
Merlin~L James, Gerald~M Smith, and James~C Wolford.
\newblock {\em Analog computer simulation of engineering systems}.
\newblock International Textbook Company, 1966.

\bibitem{johnson1963analog}
Clarence~L Johnson.
\newblock {\em Analog computer techniques}.
\newblock McGraw-Hill Book Company, Incorporated, 1963.

\bibitem{kalmar1947reduction}
Laszlo Kalmar and Janos Suranyi.
\newblock On the reduction of the decision problem.
\newblock {\em The Journal of Symbolic Logic}, 12(03):65--73, 1947.

\bibitem{kosta2011human}
Shiv~Prasad Kosta, YP~Kosta, Mukta Bhatele, YM~Dubey, Avinash Gaur, Shakti
  Kosta, Jyoti Gupta, Amit Patel, and Bhavin Patel.
\newblock Human blood liquid memristor.
\newblock {\em International Journal of Medical Engineering and Informatics},
  3(1):16--29, 2011.

\bibitem{Shahar2012}
S.~Kvatinsky, N.~Wald, G.~Satat, A.~Kolodny, U.~C. Weiser, and E.~G. Friedman.
\newblock Mrl - memristor ratioed logic.
\newblock In {\em 2012 13th International Workshop on Cellular Nanoscale
  Networks and their Applications}, pages 1--6, Aug 2012.

\bibitem{Shahar2014}
Shahar Kvatinsky, Dmitry Belousov, Slavik Liman, Guy Satat, Nimrod Wald, Eby~G.
  Friedman, Avinoam Kolodny, and Uri~C. Weiser.
\newblock {MAGIC} - memristor-aided logic.
\newblock {\em {IEEE} Trans. on Circuits and Systems}, 61-II(11):895--899,
  2014.

\bibitem{Lehtonen2014}
Eero Lehtonen, Jari Tissari, Jussi~H. Poikonen, Mika Laiho, and Lauri Koskinen.
\newblock A cellular computing architecture for parallel memristive stateful
  logic.
\newblock {\em Microelectronics Journal}, 45(11):1438--1449, 2014.

\bibitem{Linn}
E~Linn, R~Rosezin, S~Tappertzhofen, U~Böttger, and R~Waser.
\newblock Beyond von neumann—logic operations in passive crossbar arrays
  alongside memory operations.
\newblock {\em Nanotechnology}, 23(30):305205, 2012.

\bibitem{lykkebo2014mecobo}
Odd~Rune Lykkeb{\o}, Simon Harding, Gunnar Tufte, and Julian~F Miller.
\newblock Mecobo: A hardware and software platform for in materio evolution.
\newblock In {\em International Conference on Unconventional Computation and
  Natural Computation}, pages 267--279. Springer, 2014.

\bibitem{lykkebo2015investigation}
Odd~Rune Lykkeb{\o}, Stefano Nichele, and Gunnar Tufte.
\newblock An investigation of square waves for evolution in carbon nanotubes
  material.
\newblock In {\em 13th European Conference on Artificial Life}, 2015.

\bibitem{mancuso1999hydraulic}
Stefano Mancuso.
\newblock Hydraulic and electrical transmission of wound-induced signals in
  vitis vinifera.
\newblock {\em Functional Plant Biology}, 26(1):55--61, 1999.

\bibitem{mancuso1999seasonal}
Stefano Mancuso.
\newblock Seasonal dynamics of electrical impedance parameters in shoots and
  leaves related to rooting ability of olive (olea europea) cuttings.
\newblock {\em Tree physiology}, 19(2):95--101, 1999.

\bibitem{martinsen2010memristance}
{\O}~G Martinsen, S~Grimnes, CA~L{\"u}tken, and GK~Johnsen.
\newblock Memristance in human skin.
\newblock In {\em Journal of Physics: Conference Series}, volume 224, page
  012071. IOP Publishing, 2010.

\bibitem{masi2009spatiotemporal}
Elisa Masi, Marzena Ciszak, G~Stefano, Luciana Renna, Elisa Azzarello, Camila
  Pandolfi, S~Mugnai, F~Balu{\v{s}}ka, FT~Arecchi, and S~Mancuso.
\newblock Spatiotemporal dynamics of the electrical network activity in the
  root apex.
\newblock {\em Proceedings of the National Academy of Sciences},
  106(10):4048--4053, 2009.

\bibitem{massey2015computing}
MK~Massey, A~Kotsialos, F~Qaiser, DA~Zeze, C~Pearson, D~Volpati, L~Bowen, and
  MC~Petty.
\newblock Computing with carbon nanotubes: Optimization of threshold logic
  gates using disordered nanotube/polymer composites.
\newblock {\em Journal of Applied Physics}, 117(13):134903, 2015.

\bibitem{miller2014evolution}
Julian~F Miller, Simon~L Harding, and Gunnar Tufte.
\newblock Evolution-in-materio: evolving computation in materials.
\newblock {\em Evolutionary Intelligence}, 7(1):49--67, 2014.

\bibitem{mills1995kirchhoff}
J~Mills.
\newblock Kirchhoff-lukasiewicz machines, 1995.

\bibitem{mills2008nature}
Jonathan~W Mills.
\newblock The nature of the extended analog computer.
\newblock {\em Physica D: Nonlinear Phenomena}, 237(9):1235--1256, 2008.

\bibitem{morgan2016simple}
Alex~JL Morgan, David~A Barrow, Andrew Adamatzky, and Martin~M Hanczyc.
\newblock Simple fluidic digital half-adder.
\newblock {\em arXiv preprint arXiv:1602.01084}, 2016.

\bibitem{ore1960note}
Oystein Ore.
\newblock Note on hamilton circuits.
\newblock {\em The American Mathematical Monthly}, 67(1):55--55, 1960.

\bibitem{Papandroulidakis}
G.~Papandroulidakis, I.~Vourkas, N.~Vasileiadis, and G.~C. Sirakoulis.
\newblock Boolean logic operations and computing circuits based on memristors.
\newblock {\em IEEE Transactions on Circuits and Systems II: Express Briefs},
  61(12):972--976, Dec 2014.

\bibitem{perel2006ultrasound}
Mark~E Perel'man and Galina~M Rubinstein.
\newblock Ultrasound vibrations of plant cells membranes: water lift in trees,
  electrical phenomena.
\newblock {\em arXiv preprint physics/0611133}, 2006.

\bibitem{Pershin}
Yuriy~V. Pershin and Massimiliano~Di Ventra.
\newblock {Neuromorphic, Digital, and Quantum Computation With Memory Circuit
  Elements}.
\newblock {\em Proceedings of the IEEE}, 100(6):2071--2080, 2012.

\bibitem{peterson1967basic}
Gerald~R Peterson.
\newblock {\em Basic analog computation}.
\newblock Macmillan, 1967.

\bibitem{schlicht2013indole}
Markus Schlicht, Jutta Ludwig-M{\"u}ller, Christian Burbach, Dieter Volkmann,
  and Frantisek Baluska.
\newblock Indole-3-butyric acid induces lateral root formation via
  peroxisome-derived indole-3-acetic acid and nitric oxide.
\newblock {\em New Phytologist}, 200(2):473--482, 2013.

\bibitem{ITRS}
{Semiconductor Industry Association}.
\newblock {\em {International Technology Roadmap for Semiconductors (ITRS)}}.
\newblock {Semiconductor Industry Association}, 2007.

\bibitem{soroka1954analog}
Walter~W Soroka.
\newblock {\em Analog methods in computation and simulation}.
\newblock McGraw-Hill, 1954.

\bibitem{steinkellner2007flavonoids}
Siegrid Steinkellner, Venasius Lendzemo, Ingrid Langer, Peter Schweiger,
  Thanasan Khaosaad, Jean-Patrick Toussaint, and Horst Vierheilig.
\newblock Flavonoids and strigolactones in root exudates as signals in
  symbiotic and pathogenic plant-fungus interactions.
\newblock {\em Molecules}, 12(7):1290--1306, 2007.

\bibitem{stone2005phototropins}
Bethany~B Stone, C~Alex Esmon, and Emmanuel Liscum.
\newblock Phototropins, other photoreceptors, and associated signaling: the
  lead and supporting cast in the control of plant movement responses.
\newblock {\em Current topics in developmental biology}, 66:215--238, 2005.

\bibitem{strukov2008missing}
Dmitri~B Strukov, Gregory~S Snider, Duncan~R Stewart, and R~Stanley Williams.
\newblock The missing memristor found.
\newblock {\em nature}, 453(7191):80--83, 2008.

\bibitem{sugiyama2012root}
Akifumi Sugiyama and Kazufumi Yazaki.
\newblock Root exudates of legume plants and their involvement in interactions
  with soil microbes.
\newblock In {\em Secretions and exudates in biological systems}, pages 27--48.
  Springer, 2012.

\bibitem{tarabella2015hybrid}
Giuseppe Tarabella, Pasquale D'Angelo, Angelica Cifarelli, Alice Dimonte,
  Agostino Romeo, Tatiana Berzina, Victor Erokhin, and Salvatore Iannotta.
\newblock A hybrid living/organic electrochemical transistor based on the
  physarum polycephalum cell endowed with both sensing and memristive
  properties.
\newblock {\em Chemical Science}, 6(5):2859--2868, 2015.

\bibitem{trewavas2005green}
Anthony Trewavas.
\newblock Green plants as intelligent organisms.
\newblock {\em Trends in plant science}, 10(9):413--419, 2005.

\bibitem{trewavas2009plant}
Anthony Trewavas.
\newblock What is plant behaviour?
\newblock {\em Plant, cell \& environment}, 32(6):606--616, 2009.

\bibitem{trewavas2011ubiquity}
Anthony~J Trewavas and Franti{\v{s}}ek Balu{\v{s}}ka.
\newblock The ubiquity of consciousness.
\newblock {\em EMBO reports}, 12(12):1221--1225, 2011.

\bibitem{volkov2006electrophysiology}
Alexander~G Volkov.
\newblock Electrophysiology and phototropism.
\newblock In {\em Communication in Plants}, pages 351--367. Springer, 2006.

\bibitem{volkov2006plants}
Alexander~G Volkov and Don Rufus~A Ranatunga.
\newblock Plants as environmental biosensors.
\newblock {\em Plant signaling \& behavior}, 1(3):105--115, 2006.

\bibitem{volkov2014memristors}
Alexander~G Volkov, Clayton Tucket, Jada Reedus, Maya~I Volkova, Vladislav~S
  Markin, and Leon Chua.
\newblock Memristors in plants.
\newblock {\em Plant signaling \& behavior}, 9(3):e28152, 2014.

\bibitem{Vourkas}
I.~Vourkas and G.~Ch. Sirakoulis.
\newblock {Emerging Memristor-Based Logic Circuit Design Approaches: A Review}.
\newblock {\em IEEE Circuits and Systems Magazine}, 16(3):15--30, thirdquarter
  2016.

\bibitem{VourkasS14}
Ioannis Vourkas and Georgios~Ch. Sirakoulis.
\newblock Memristor-based combinational circuits: A design methodology for
  encoders/decoders.
\newblock {\em Microelectronics Journal}, 45(1):59--70, 2014.

\bibitem{Vourkasbook}
Ioannis Vourkas and Georgios~Ch Sirakoulis.
\newblock {\em {Memristor-based nanoelectronic computing circuits and
  architectures}}.
\newblock Emergence complexity and computation. Springer, Cham, 2016.

\bibitem{weyrick1969fundamentals}
Robert~C Weyrick.
\newblock {\em Fundamentals of analog computers}.
\newblock Prentice Hall, 1969.

\bibitem{xu2013improved}
Weifeng Xu, Guochang Ding, Ken Yokawa, Franti{\v{s}}ek Balu{\v{s}}ka, Qian-Feng
  Li, Yinggao Liu, Weiming Shi, Jiansheng Liang, and Jianhua Zhang.
\newblock An improved agar-plate method for studying root growth and response
  of arabidopsis thaliana.
\newblock {\em Scientific reports}, 3:1273, 2013.

\bibitem{yokawa2014binary}
Ken Yokawa and Frantisek Baluska.
\newblock Binary decisions in maize root behavior: Y-maze system as tool for
  unconventional computation in plants.
\newblock {\em IJUC}, 10(5-6):381--390, 2014.

\bibitem{yokawa2011illumination}
Ken Yokawa, Tomoko Kagenishi, Tomonori Kawano, Stefano Mancuso, and
  Franti{\v{s}}ek Balu{\v{s}}ka.
\newblock Illumination of arabidopsis roots induces immediate burst of ros
  production.
\newblock {\em Plant signaling \& behavior}, 6(10):1460--1464, 2011.

\end{thebibliography}

\printindex

\end{document}